\documentclass[jcp,twocolumn,longbibliography,superscriptaddress]{revtex4-2}
\usepackage{graphicx}
\usepackage[utf8]{inputenc}
\usepackage{amsmath}
\usepackage{verbatim}
\usepackage{amssymb}
\usepackage{listings}
\usepackage{color}
\usepackage{url}
\usepackage{hyperref}
\usepackage{amsmath}
\usepackage{mathtools}
\usepackage{xspace}
\usepackage{float}
\usepackage{multirow}
\usepackage[caption=false]{subfig}

\usepackage{ifxetex}
\ifxetex
\usepackage{fontspec}
\else
\usepackage[T1]{fontenc}
\usepackage[utf8]{inputenc}
\fi

\newcommand{\tmix}[1][]{t_{\text{mix}}}
\newcommand{\tcorr}[1][]{t_{\text{corr}}}
\newcommand{\pit}[1]{\pi^{\{#1\}}}

\newcommand{\ultranaive}{naive\xspace}

\newcommand{\tsim}{\tau_{\text{sim}}}

\newcommand{\Area}{V}
\newcommand{\MinArea}{V_0}
\newcommand{\swell}{homothetic\xspace}
\newcommand{\np}{non-periodic\xspace}
\newcommand{\HD}{\texttt{HistoricDisks}\xspace}
\newcommand{\piOne}{\pi^{(1)}}

\newcommand{\nunitwallx}{\nhat_{\text{wall}}^{\pm \ehatvec_x} }
\newcommand{\nwallx}{n_{\text{wall}}^{\pm x} }
\newcommand{\nwallxy}{n_{\text{wall}}^{\pm \ehatvec_x \pm \ehatvec_y}}
\newcommand{\npair}{n_{\text{pair}} }
\newcommand{\vperp}{v^\perp}
\newcommand{\vperpwall}{\vperp_{\text{wall}}}
\newcommand{\vrelperp}{\Delta v^{\perp}} 
\newcommand{\vrelperppair}{\vrelperp_{ij}}
\newcommand{\aspect}[2]{\alpha\! =\! (#1 \! :  \! #2)}
\newcommand{\aspectclean}[2]{ (#1 \! :  \! #2)}
\newcommand{\RR}{R_v}
\newcommand{\scalfac}{\gamma}

\newcommand{\epsswell}{\epsilon_\alpha}
\newcommand{\rhox}{\rho_x}
\newcommand{\rhoy}{\rho_y}
\newcommand{\delxmin}{\Delta x^{\min}}
\newcommand{\delxminij}{\delxmin_{ij}}
\newcommand{\delymin}{\Delta y^{\min}}
\newcommand{\wallcolls}[2][]{w:(#2 #1)}
\newcommand{\paircolls}{p:(ij)}
\newcommand{\simasymp}{\stackrel{N  \to  \infty}{\longrightarrow}}
\newcommand{\GDot}[2][]{\frac{\diff}{\diff t}G_{#2}^{\text{#1}}}

\newcommand\subfig[2]{{Fig.~\ref{#1}{#2}}}
\newcommand\subcap[1]{{(#1):}}
\newcommand\subcaptwo[2]{{(#1) and (#2):}}

\newcommand{\SET}[1]{\{#1\}}

\newcommand{\eq}[1]{Eq.~\eqref{#1}}
\newcommand{\eqtwo}[2]{Eqs~\eqref{#1} and~\eqref{#2}}

\newcommand{\Eq}[1]{Eq.~\eqref{#1}}
\newcommand{\fig}[1]{Fig.~\ref{#1}}

\newcommand{\quot}[1]{``#1''}
\newcommand{\tab}[1]{Table~\ref{#1}} 
 
\newcommand{\app}[1]{Appendix~\ref{#1}} 
\newcommand{\sect}[1]{Section~\ref{#1}} 
\newcommand{\subsect}[1]{Subsection~\ref{#1}}

\newcommand{\cf}{\textrm{cf}}
\newcommand{\etc}{\textrm{etc.}}
\newcommand{\etal}{\textrm{et al.}}

\newcommand{\vs}{\textrm{vs.}}

\newcommand{\OCAL}{\mathcal{O}}  
\newcommand{\expb}[1]{\exp \glb #1 \grb} 
\newcommand{\sina}[2][]{\sin^{#1} \! \gla #2 \gra}  
\newcommand{\cosa}[2][]{\cos^{#1} \! \gla #2 \gra}  

\newcommand{\loga}[2][]{\log^{#1}\! \gla #2 \gra}  
\newcommand{\logc}[2][]{\log^{#1} \glc #2 \grc}  

\newcommand{\prob}{\mathbb{P}}

\newcommand{\gla}{\,}  
\newcommand{\gra}{}  
\newcommand{\glb}{\left(}  
\newcommand{\grb}{\right)}  
\newcommand{\glc}{\left[}  
\newcommand{\grc}{\right]}  

\newcommand{\PLUSPLUS}{+ \dots +}

\newcommand{\TO}{,\ldots,}
\newcommand{\VEC}[1]{\mathbf{#1}}
\newcommand{\ehatvec}{\hat{\VEC{e}}}

\newcommand{\vvec}{\VEC{v}}

\newcommand{\xvec}{\VEC{x}}

\newcommand{\ghat}{\hat{g}}

\newcommand{\nhat}{\hat{n}}

\newcommand{\scal}[2]{(#1 \pmb{\cdot} #2)}
\newcommand{\mean}[1]{\left\langle #1 \right\rangle}
\newcommand{\half}{\frac{1}{2}}

\newcommand\bigOb[1]{\ensuremath{\OCAL\glb #1 \grb}}

\newcommand\diff[1]{\mathrm{d}#1}

\newcommand{\fpn}[2]{\ensuremath{#1 \! \times \! 10^{#2}}}

\begin{document}

\title{Hard-disk computer simulations---a historic perspective}
\author{Botao Li}
\affiliation{Laboratoire de Physique de l’Ecole normale sup\'erieure, ENS,
Universit\'e PSL, CNRS, Sorbonne Universit\'e, Universit\'e de Paris Cit\'e, 
Paris,
France}
\author{Yoshihiko Nishikawa}
\affiliation{Graduate School of Information Sciences, Tohoku University, Sendai
980-8579, Japan}

\author{Philipp H\"ollmer}
\affiliation{Physikalisches Institut and Bethe Center for Theoretical Physics,
University of Bonn, Nussallee 12, 53115 Bonn, Germany}
\author{Louis Carillo}
\affiliation{Laboratoire de Physique de l’Ecole normale sup\'erieure, ENS,
Universit\'e PSL, CNRS, Sorbonne Universit\'e, Universit\'e de Paris Cit\'e, 
Paris,
France}
\author{A. C. Maggs}
\affiliation{CNRS Gulliver, ESPCI Paris, Universit\'e PSL, 10 rue Vauquelin,
75005 Paris, France.}
\author{Werner Krauth}
\email{werner.krauth@ens.fr}
\affiliation{Laboratoire de Physique de l’Ecole normale sup\'erieure, ENS,
Universit\'e PSL, CNRS, Sorbonne Universit\'e, Universit\'e de Paris Cit\'e, 
Paris,
France}

\date{\today}

\begin{abstract}
We  discuss historic pressure computations for the hard-disk model performed 
since 1953, and compare them to results that we obtain with a powerful 
event-chain Monte Carlo and a massively parallel Metropolis algorithm. Like 
other simple models in the sciences, such as the Drosophila model of biology, 
the hard-disk model has needed monumental effort to be understood. In 
particular, 
we argue that the difficulty of estimating the pressure has not been fully 
realized in the decades-long controversy over the hard-disk phase-transition 
scenario. We present the physics of the hard-disk model, the definition of the 
pressure and its unbiased estimators, several of which are new. We further treat 
different sampling algorithms and crucial criteria for bounding mixing 
times in the absence of analytical predictions. Our definite results for the 
pressure, for up to one million disks, may serve as benchmarks for future 
sampling algorithms.  A synopsis of hard-disk pressure data as well as different 
versions of the sampling algorithms and pressure estimators are made available 
in an open-source repository.
\end{abstract}

\maketitle%


\section{Introduction}
\label{sec:Introduction}
In fundamental physics, the most detailed descriptions of 
physical reality 
are not always the
best. In our quantum-mechanical world, many
phenomena are
indeed 
described classically,
without the elaborate machinery of
wavefunctions and 
density matrices.
The  exact thermodynamic singularities of
helium, a quantum liquid, at the famous lambda transition~\cite{Kapitza1938}
are for example obtained by a seemingly unrelated model of 
classical 
two-component
spins~\cite{Vaks1966phase,CampostriniHasenbuschVicari2006,Hasenbusch_Clock2019,
Chester2020} on a three-dimensional lattice rather than 
by some quantum-mechanical  representation of all atoms in
the continuum~\cite{Pollock1987}.
Renormalization-group
theory~\cite{Wilson1983} guarantees that the
simple classical spin model 
exactly describes experiments 
in the quantum liquid~\cite{Lipa1996,Lipa2003}. 
The universality of simple models is
also found in other sciences. In biology,  the study of the fruit
fly Drosophila has gradually evolved from a subject of entomology, the science 
of insects,
to  a parable for higher animals, where it allows one  to 
appreciate gene
damage~\cite{Mueller1946} from radiation. In recent
decades, it was moreover understood that many of the  genes of the 
Drosophila have exactly the same function as genes in vertebrates, including
humans. In physics as in biology, \quot{(t)his remarkable conservation came as a
big surprise. It had been neither predicted nor 
expected}~\cite{Nuesslein1995},
to cite a Nobel-prize winner.

Paradoxically, even  simple models in  science, those stripped to their 
bare
bones, can take  monumental effort and decades of research to be understood
fully. This is the case for the Drosophila fly that entered 
research
laboratories around 1905~\cite{Kohler1994Lords}, and then gradually turned into
a model organism~\cite{Dietrich2014}.
It is also what happened to
the simplest of all particle models, the hard-disk model, which is the
focus of the present work. The model consists of $N$ 
identical classical disks with
positions inside a box and with velocities.
Disks fly in straight-line trajectories, except when they 
collide with each 
other or with an enclosing wall.
The elementary 
collision rules are borrowed from billiards.
The hard-disk model caricatures 
two-dimensional fluids: It lacks the
explicit interparticle attractions of more 
realistic
descriptions, yet it shows almost all the interesting properties 
of
general particle systems. Moreover, its observed phase 
behavior~\cite{Alder1962} was 
understood 
in terms of topological phase transitions, just like  
classical continuous 
spin models in two dimensions~\cite{KosterlitzThouless1973}. The 
interpretation common to both cases was 
unsuspected from the conventional Landau theory of phase transitions.

Only few characteristics of the hard-disk model are known from rigorous 
mathematics. The first was proven  by Daniel Bernoulli in 
1738~\cite{Bernoulli1738}, namely that the temperature, which is linked to the 
mean-square velocity of the disks, 
plays no role in their spatial distribution.
It was also proven~\cite{Sinai1970,Simanyi2003} that 
the 
hard-disk model, as a dynamical system with positions and 
velocities evolving under Newton's laws, can be described 
statistically 
with positions that all have equal 
statistical weight. It is furthermore shown rigorously that, at small finite 
density, the 
model is fluid~\cite{LebowitzPenrose1964,Helmuth2022}, justifying
analytic expansions  developed in the  $19$th century~\cite{BoltzmannBook} to link 
dilute hard disks to the ideal gas of non-interacting particles.
For high densities, 
it was established that at close packing,
the hard-disk model forms a hexagonal crystal~\cite{Fejes1940}. 
However, for all densities 
below close packing, this crystalline structure is 
destroyed~\cite{Richthammer2007,Richthammer2016} by 
long-wavelength fluctuations. The invention  of simulation 
methods, and their application to this very model of hard disks ever since the 
1950s, was meant to overcome the scarcity of analytical results.

\begin{figure}[htb]
\centering
\includegraphics[width=\columnwidth]
{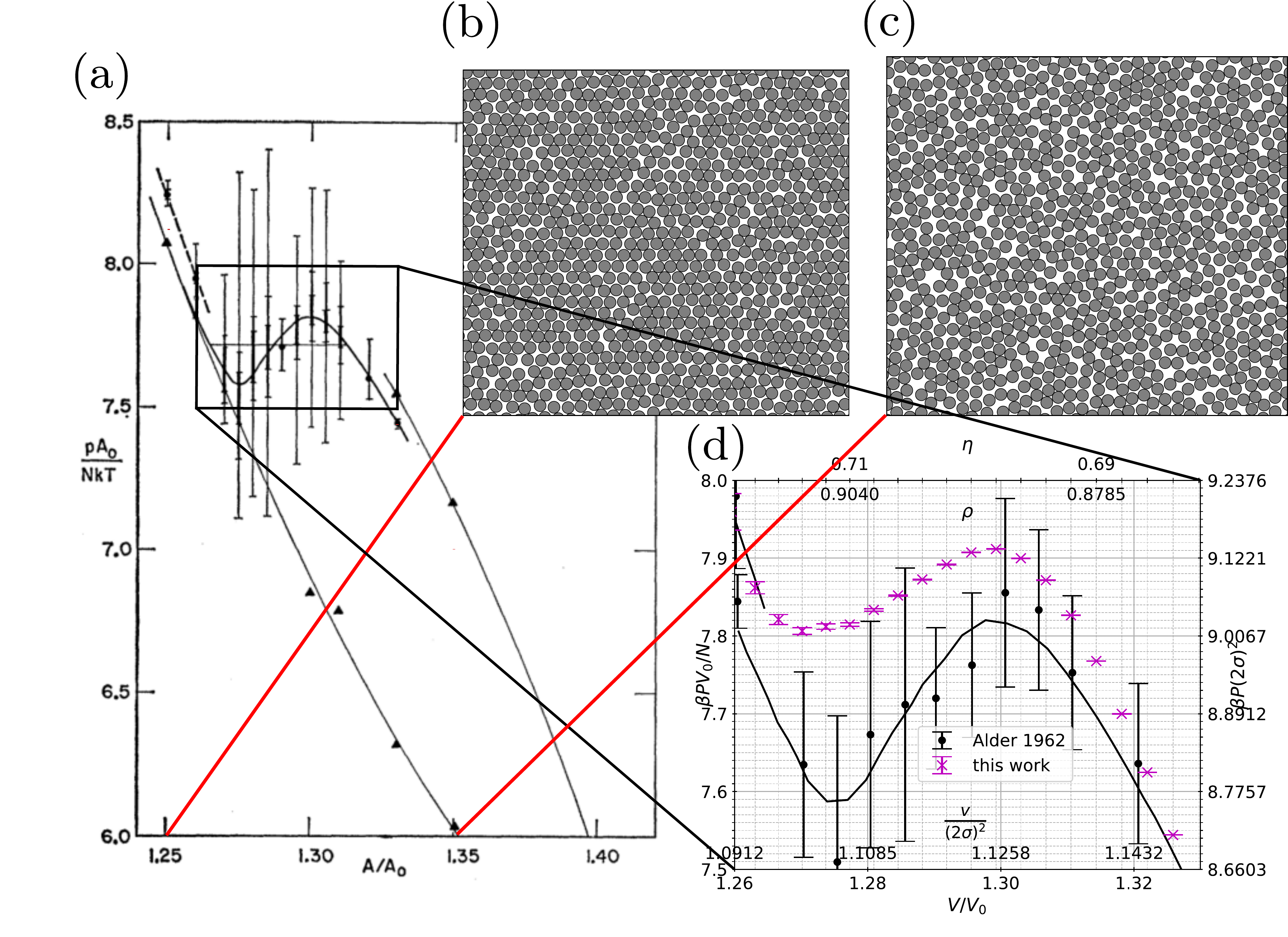}
\caption{Hard disks in a periodic box.
\subcap{a} Equation of state (pressure \vs\ volume) computed in 1962
by Alder and Wainwright~\cite{Alder1962}
\subcaptwo{b}{c} Samples of $870$ disks at densities $\eta = 0.726$ and 
$\eta =
0.672$. \subcap{d} Pressures for $N=870$ from (a) compared with ECMC 
results (this work) for aspect 
ratio  $\aspect{9}{8 \sqrt{3}/2}$ 
(\cf\ \fig{fig:Psi6Dependence} for an analysis 
of convergence behavior).}
\label{fig:AlderWainwrightRevisited}
\end{figure}

With its stripped-down interactions, 
the hard-disk model is indeed simple. In particular, the model lacks
attractive forces that would pull the disks together. It is for this reason 
that, for a
long time, hard disks and hard spheres (in three dimensions) were considered 
too simple to show a 
phase
transition from an disordered fluid to a solid~\cite{KirkwoodMonroe1941,
Battimelli2018}. In two dimensions, furthermore, ordered phases were
expected not to exist for theoretical reasons that were considered
sufficiently solid to formally exclude any 
transition~\cite{Peierls1935}. Initial
computer simulations, in 1953, in the same publication that introduced the
Metropolis algorithm~\cite{Metropolis1953}, accordingly found
that \quot{(t)here is no indication of a phase transition}. It thus came as an
enormous surprise when, in 1962, Alder and Wainwright~\cite{Alder1962} 
identified a loop in the equation of state (see 
\fig{fig:AlderWainwrightRevisited}),
suggesting~\cite{MayerWood1965} a phase transition between a 
fluid at low density
and a solid (that
was not supposed, at the time, to exist) at high density.
This laid the ground work
for the Kosterlitz--Thouless--Halperin--Nelson--Young theory 
of melting in two-dimensional particle 
systems~\cite{Kosterlitz1974,HalperinNelson1978,Young1979VectorCoulomb}.
Even after this important conceptual advance,
the phase behavior of the hard-disk model 
remained controversial for another fifty years, until an 
analysis~\cite{Bernard2011}
based on the powerful event-chain Monte Carlo 
(ECMC) algorithm~\cite{Bernard2009} showed
that hard disks melt 
through a first-order fluid--hexatic phase transition combined
with a Kosterlitz--Thouless transition between the hexatic and
the solid, thus proposing a new scenario.

In this work, we discuss the hard-disk model in a computational and historical 
context, concentrating on the pressure as a function of the volume. The reason 
for this restriction of scope on pressures rather than on phases is
that the aforementioned fifty-year controversy was rooted in difficulties in 
computing the pressure.
After a 
short introduction to the physics of the hard-disk model, we review the
thermodynamic and kinematic pressure
definitions, and show that they are perfectly equivalent 
even for finite systems.
Nevertheless, the 
pressure may be anisotropic and 
depend on the shape of the system, rather than being only a function of system 
volume. 
We 
discuss pressure estimators, with a focus on
those that are 
unbiased
at finite $N$ and convenient to use.
We furthermore clarify 
that different sampling algorithms (molecular dynamics, reversible and 
non-reversible MCMC) all rigorously sample  the same equal-weight Boltzmann 
distribution of positions although 
the time scales for convergence can differ widely even for 
local algorithms, and can reach years and even decades of computer time for 
moderate system sizes. This was not understood in many important historical 
contributions.
With all 
this material in hand, we  confront past results with massive 
computations performed for this work, thus providing
definite high-precision pressure estimates for the hard-disk model that may 
serve as benchmarks 
for the future. 
With its rich 
phenomenology and its intractable mathematics, 
this simple model has become 
the 
Drosophila for particle systems and for two-dimensional phase transitions. 
It has served as a parable for difficult computing problems and 
as a platform for development of MCMC and of molecular dynamics. 
This fascinating model has yet to be fully understood. We aim at providing 
a solid base for future work.

The plan of this work is as follows. In \sect{sec:HardDiskBasics}, we discuss 
the fundamentals of the hard-disk model, from the definitions of densities and 
volumes to an overview of its physical properties. In 
\sect{sec:Algorithms}, we discuss sampling algorithms (molecular dynamics  and 
Markov-chain Monte Carlo (MCMC)) and pressure estimators for this model, 
concentrating 
on new developments. In \sect{sec:EquationsOfState}, we digitize, discuss, 
and make available numerical computations of the equation of state performed 
since the paper by Metropolis \etal, in 1953, and superpose 
them with 
large-scale computations done for this work. \sect{sec:Conclusion} contains our 
conclusions. We also provide information on statistical analysis 
(\app{app:Statistics}) and introduce to the \HD\ open-source 
software package, which collects pressure data since 1953, and implements 
sampling algorithms and pressure estimators (\app{app:DataCodes}) that were used
in this work.

\section{Hard-disk model}
\label{sec:HardDiskBasics}

The hard-disk model consists of $N$ impenetrable,
identical, classical disks of radius $\sigma$ and mass $m$, that 
are perfectly elastic. Collisions are 
instantaneous; they cause no deformations and induce no rotations. 
Pair collisions conserve momentum and energy. Disks evolve in a 
fixed rectangular box of sides $L_x$ and $L_y$ (specified through the aspect 
ratio $\aspect{L_x}{L_y}$), which may have periodic boundary conditions 
(\quot{periodic} box), or else hard-wall boundary conditions 
(\quot{\np} box).
The two-dimensional volume (area) is $V = L_x \times L_y$. 
In the $NVT$ ensemble that we 
consider here, $N$, $L_x$, and $L_y$  are fixed. For the hard-disk system, the 
microcanonical ensemble (of constant energy $E$) and the 
canonical ensemble (of constant temperature $T$) are almost equivalent, and we 
connect the two throughout this work. 
In other ensembles, the box can be of varying 
dimensions~\cite{Lee1992,Andersen1980,Parrinello1981}, and $N$ might 
vary~\cite{Rowley1975}. 
The disk $i$ is described 
by the position of its center $\xvec_i = (x_i,y_i)$, and possibly 
by a 
velocity $\vvec_i = (v_{x,i}, v_{y,i})$. We denote the set of positions and 
velocities by $\xvec = \SET{\xvec_1 \TO \xvec_N}$   and $\vvec = \SET{\vvec_1 
\TO \vvec_N}$, respectively.

\subsection{Basic definitions and properties}

We now discuss additional characteristics of the hard-disk model
and define its Newtonian dynamics.
Furthermore, we summarize the 
physics 
of two-dimensional particle systems.

\subsubsection{System definitions, basic properties}

\begin{figure}[htb]
\centering
\includegraphics[width=0.8\columnwidth]
{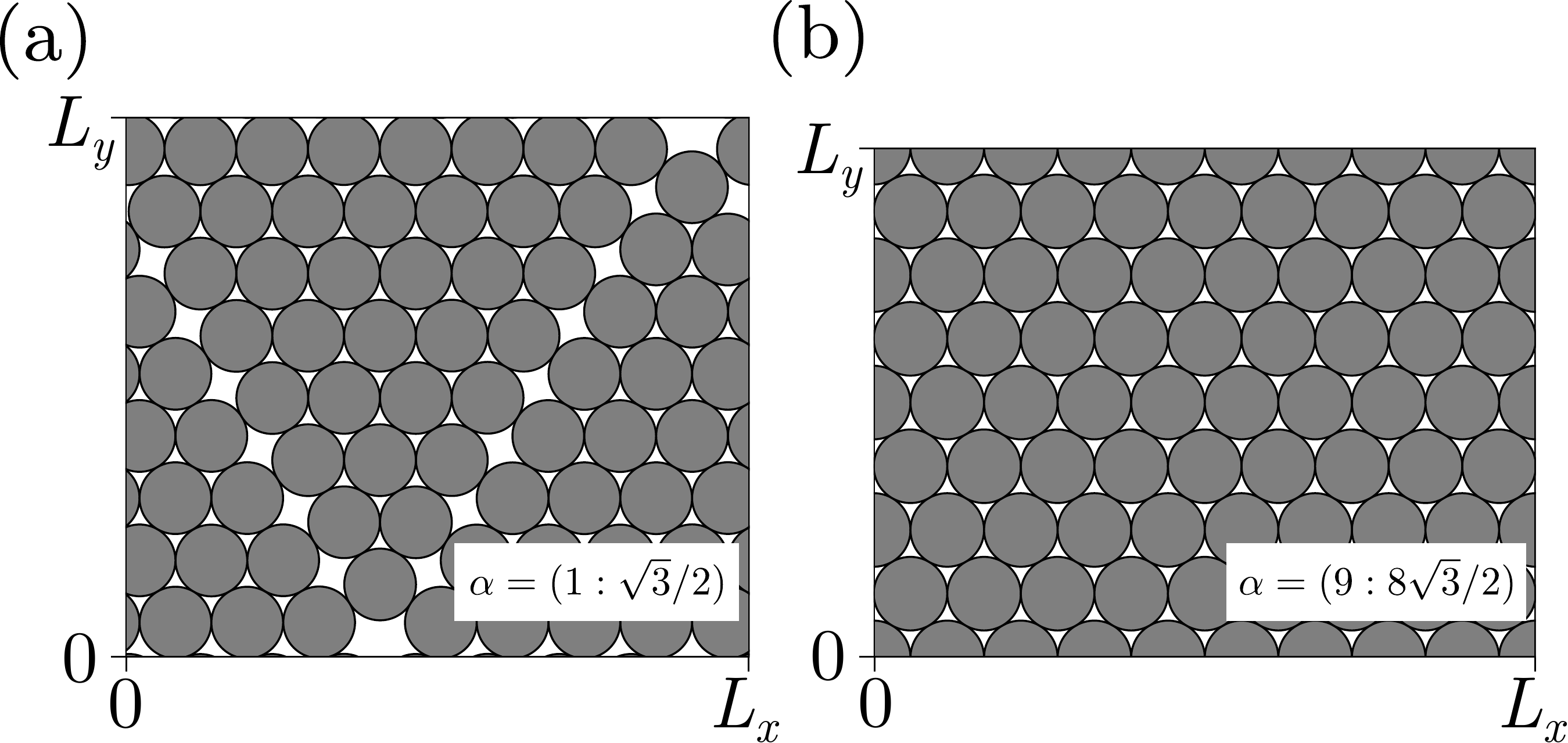}
\caption{Packings for $N=72$ disks for different
aspect ratios
$\alpha$. \subcap{a} Periodic box with $\alpha = (1: \sqrt{3}/2)$, with
conjectured optimal packing. 
\subcap{b} Periodic box with $\alpha = (9 : 8 \sqrt{3}/2)$ at the close-packing
density $\eta = \pi / (2\sqrt{3})$. }
\label{fig:SolidInit}
\end{figure}

In the limit $N \to \infty$, the properties of a particle system with 
short-range interactions are independent of the boundary conditions. At finite 
$N$, in contrast, the bulk part of the free energy (that scales as $V$) cannot 
be separated from the surface term (that, in two dimensions, scales as 
$\sqrt{V}$). For example, a close-packed crystal of $N=72 = 8 \times 9$ disks 
can 
fit into a periodic box of aspect ratio  $\aspect{9}{8 \sqrt{3}/2}$, whereas the 
maximum density for the same $N$ in a box with $\aspect{1}{\sqrt{3}/2}$ is 
slightly smaller (see \fig{fig:SolidInit}). 
Evidently, the pressure in a box containing a finite number of disks depends on 
its aspect ratio. 

The following conventions for the volume or its inverse, the density, have been 
commonly
used in the literature. The first is the volume $V = L_x 
\times L_y$ 
normalized by that of a perfect
hexagonal 
packing $V_0 = N (2 \sigma)^2 \sqrt{3} / 2$ (as in \subfig{fig:SolidInit}{b}). 
Second is the covering density $\eta$, the 
total volume of all disks normalized by the box volume $V$. Third is the 
reduced density $\rho$, the number $N$ of disks of radius $\half$ divided 
by the volume and, finally, the inverse normalized density $\nu /(2 \sigma)^2$ 
with $\nu=1/\rho$. These quantities are related as follows:
\begin{equation}
\begin{aligned}
\frac{V}{V_0} &= \frac{\pi}{2 \sqrt{3}} \frac{1}{\eta} = \frac{2}{\sqrt{3}}
\frac{\nu}{(2 \sigma) ^2} \ge 1, \\\
\eta &= \frac{\pi}{2\sqrt{3}} \frac{\MinArea}{\Area}= \frac{N}{V} \pi
\sigma^2
\le 0.907,
\\
\rho &= \eta \frac{4}{\pi} = \frac{\MinArea}{\Area} \frac{2}{\sqrt{3}} =
\frac{(2 \sigma) ^2}{\nu} \le 1.155, \\
\frac{\nu}{(2 \sigma) ^2 } & = \frac{\sqrt{3}}{2} \frac{\Area}{\MinArea} =
\frac{\pi}{4} \frac{1}{\eta} = \frac{1}{\rho}  \ge 0.866.
\end{aligned}
\label{equ:DensityVolume}
\end{equation}
We will re-plot published equations of state with all four units, 
thus simplifying the comparison of data.

In the hard-disk model, all configurations have zero potential energy, and the 
Newtonian time evolution conserves the kinetic energy.
Pairs of disks collide such that, 
in their center-of-mass coordinate system, they
rebound from their line of contact
with conserved parallel and reversed perpendicular 
velocities~\cite{SMAC}. At a wall collision, 
the velocity component of a disk parallel to the wall remains the same while 
the 
perpendicular
velocity
$\vperp_{\text{wall}}$ is reversed. 
When, at the initial time $t=0$, all
velocities are rescaled by a factor $\scalfac$, 
the entire trajectory transforms as 
\begin{equation}
\SET{\xvec_1(t) \TO \xvec_N(t) } \xrightarrow{\vvec_i \to \scalfac \vvec_i 
\forall i
} \SET{\xvec_1(\tfrac{t}{\scalfac}) \TO \xvec_N(\tfrac{t}{\scalfac}) }.
\label{equ:BernoulliScaling}
\end{equation}
This property of hard-sphere models was already noticed by Daniel 
Bernoulli~\cite{Bernoulli1738}. 

Statistically, during the Newtonian hard-disk 
time evolution, 
the sets of 
positions and of velocities are mutually independent.
All positions $\xvec$ that correspond to $N$-disk configurations 
have the same statistical weight $\pi$
with a Cartesian density measure
and, 
in a \np box,
the velocities are randomly distributed 
on the surface of the hypersphere in $2N$-dimensional space with radius 
$\RR = \sqrt{2E/m}$, where 
$E$ is the conserved (microcanonical) energy. The measure in the 
$4N$-dimensional sample space 
of positions and velocities is thus 
\begin{equation}
  \diff \pi \propto \Theta(\xvec) \diff^N\xvec\; \diff^N\vvec\, \delta \glb E 
-\sum_i 
m \vvec_i^2/2 \grb,  
  \label{equ:hypersphere}
\end{equation}
where $\Theta(\xvec)=1$ for positions that correspond to hard-disk 
configurations and $\Theta(\xvec)=0$ otherwise.
In a periodic box, in addition, the two 
components of the total velocity and the position of the center of mass 
in the rest frame are conserved.
For large $N$, where the ensembles are equivalent, the 
microcanonical energy per disk corresponds to 
$E/N = k_\text{B}T = 1/\beta$, where $k_{\text{B}}$ is the Boltzmann 
constant and $T$ the temperature of the canonical ensemble.
The spatial part of the measure of \eq{equ:hypersphere} remains unchanged, and 
the uniform 
velocity distribution on the surface of the hypersphere in $2N$ 
dimensions 
implies  that the one-particle, marginal distribution of velocity components 
becomes Gaussian:
\begin{equation}
\pi(v_{i,x})   \propto  \expb{-\beta m v_{i,x}^2/2} \quad (N \to \infty) 
\label{equ:GaussianLargeN}
\end{equation}
(and analogously for $v_{i,y}$, see~\cite[Sect. 2.1.1]{SMAC}). 

The probability distribution of the velocity perpendicular to a wall 
$\vperp_{\text{wall}}$ (essentially the 
histogram of momentum transfers with the walls at the discrete wall-collision 
times) differs from \eq{equ:hypersphere}.
For $N\to \infty$, this distribution is biased by a factor $|\vperpwall|$
with respect to the Maxwell distribution:
\begin{equation}
\pi(|\vperpwall|) \propto  | \vperpwall | \expb{-\beta m (\vperpwall) ^2/2}, 
\end{equation}
which has been described through the \quot{Maxwell boundary condition}
(see~\cite[Sect. 2.3.1]{SMAC}). 
For finite $N$, the same biasing factor appears. The distribution of the 
velocity perpendicular to a wall
is derived from 
the surface element on the hypersphere of radius $\RR = \sqrt{v_1^2 
\PLUSPLUS 
v_n^2}$
in $n=2N$ dimensions:
\begin{equation}
\diff \Omega = \RR^{n-1}\sina[n-2]{\phi_1} \sina[n-3]{\phi_2} \dots \sina 
{\phi_{n-2} }
\diff \phi_1 \dots \diff \phi_{n-1}, 
\end{equation}
where $\phi_1 \TO \phi_{n-2} \in [0, \pi]$ and $\phi_{n-1} \in [0, 2\pi]$, and 
where only $v_1 = \RR \cosa{\phi_1}$ is expressed in terms of a single angle. It 
is thus convenient to identify $v_1$ with $\vperpwall$. The radius $\RR$ of the 
hypersphere at the microcanonical energy $E = m \RR^2/2$ is  related to the 
inverse temperature  in the canonical ensemble as $\RR^2  = 2N/(m\beta)$. With 
the integrals 
\begin{equation}
\begin{aligned}
A &= \int_0^\pi \diff \phi_1 | \cosa{\phi_1} |
\sina[n-2]{\phi_1} = \frac{2}{n-1},  
\\
B &= \int_0^\pi \diff \phi_1  
\sina[n-2]{\phi_1} = \sqrt{\pi} \frac{\Gamma[(n-1)/2]}{ \Gamma(n/2)},
\end{aligned}
\end{equation}
this yields:
\begin{subequations}
\begin{align}
\mean{\frac{1}{|\vperpwall|}} & = 
\frac{1}{\RR}  
\frac{B}{A} = 
\frac{\sqrt{\pi}}{\RR}
\frac{\Gamma(N + \half)}{ \Gamma(N)} \simasymp \sqrt{\frac{\pi m \beta}{2}},
\label{equ:MaxwellBoundaryA}
\\
\mean{|\vperpwall|} & = \RR \frac{ B}{ 2 N A } = 
\frac{\RR \sqrt{\pi}}{2N} \frac{\Gamma (N + \half)}{ \Gamma(N)} \simasymp
\sqrt{\frac{\pi}{2 m \beta}},
\label{equ:MaxwellBoundaryB}
\end{align}
\label{equ:MaxwellBoundary}%
\end{subequations}
where in the limit $N \to \infty$ the 
ratio of the $\Gamma $ functions approaches $\sqrt{N}$.
The relative perpendicular velocities $\vrelperppair$ (the 
projection of the 
relative velocity $\vvec_i - \vvec_j$ perpendicular to the interface separating 
disks $i$ and $j$ at their collision) is, similarly: 
\begin{subequations}
\begin{align}
\mean{\frac{1}{|\vrelperppair|}} & = 
\frac{\sqrt{2\pi}}{\RR}
\frac{\Gamma(N + \half)}{ \Gamma(N)} \simasymp \sqrt{\pi m \beta}, 
\label{equ:MaxwellPairA}
\\
\mean{|\vrelperppair|} & = 
\frac{\RR \sqrt{\pi}}{\sqrt{2} N} \frac{\Gamma (N + \half)}{ \Gamma(N)}
\simasymp \sqrt{\frac{\pi}{ m \beta}}.
\label{equ:MaxwellPairB}
\end{align}
\label{equ:MaxwellPair}%
\end{subequations}

\subsubsection{Pair correlations, entropic phase transition}
\label{subsec:EntropicPhaseTransition}

In the hard-disk model, the Boltzmann weights are the same for all 
sets of disk positions, since there are no explicitly varying interactions. 
In consequence,  the 
two possible fluid phases (namely the gas and the liquid) are confounded. A 
purely entropic
\quot{depletion} interaction~\cite{Asakura1954} between disks nevertheless 
arises from the presence of other disks, effectively driving phase 
transitions. The three phases of the hard-disk model are fluid (with 
exponential decays of the orientational and positional
correlation functions), hexatic (with an algebraic 
decay of orientational and exponential decay of positional 
correlations), and solid (with long-range orientational correlations 
and an algebraic decay of positional correlations). The hexatic and solid 
phases have only been identified through numerical simulations, and 
mathematical proofs of their existence are still lacking. 

\subsection{Hard-disk thermodynamics}

In statistical mechanics, a homogeneous system (composed of, say, $N$ particles 
in a fixed box) is described by an equation of state connecting two quantities, 
as for example the volume and the pressure. 
When for some volumes, a homogeneous phase may not exist, two (or 
exceptionally three) phases may 
coexist. We now link the definitions of 
the pressure from the thermodynamic and kinematic viewpoints and then discuss 
phase coexistence in finite systems. 

\subsubsection{Pressure, thermodynamic and kinematic definitions}
\label{subsubsec:PressurePartitionFunction}

In statistical mechanics, the pressure $P$ is given by the
change of the free energy with the system volume: 
\begin{equation}
\beta P = \frac{\partial \loga{Z}}{\partial V}
\stackrel{V'\to V}{=} \frac{1 }{V-V'} \frac{Z - Z'}{Z}.
\label{equ:PressureElimination}
\end{equation} 
with $Z$ the partition function and 
$Z' \equiv Z(V')$.
For hard disks and 
related models, the rightmost fraction in
\eq{equ:PressureElimination} expresses the probability that a sample in the 
original box of volume $V$ is eliminated in the box of reduced volume $V' < 
V$. In  rift-pressure 
estimators~\cite{Michel2014JCP}, the volume $V$ of  
an $L_x \times L_y$ box is reduced by removing an
infinitesimal vertical or horizontal slab (a \quot{rift}), yielding the 
components 
$P_x$ and $P_y$
of the pressure. Rifts can be placed anywhere in the box, 
and one may even 
average any given hard-disk sample over all vertical or horizontal rift 
positions (see \subsect{subsec:PressureComputations}). 
We will also discuss \swell pressure 
estimators, where all box dimensions and disk coordinates are scaled down by a
common factor. Used for decades, they estimate the pressure $P = (P_x + 
P_y)/2$.

Besides the thermodynamic definition of the pressure, one can also define the
kinematic pressure as the exchange of momentum with the enclosing walls. 
However, thermodynamic and kinematic  pressures are rigorously 
identical already at finite $N$, and the corresponding estimators can be 
transformed into each another. This is also true for the pressure 
estimators built on the virial formalism that we also discuss.

\begin{figure}[htb]
\centering
\includegraphics[width=\columnwidth]{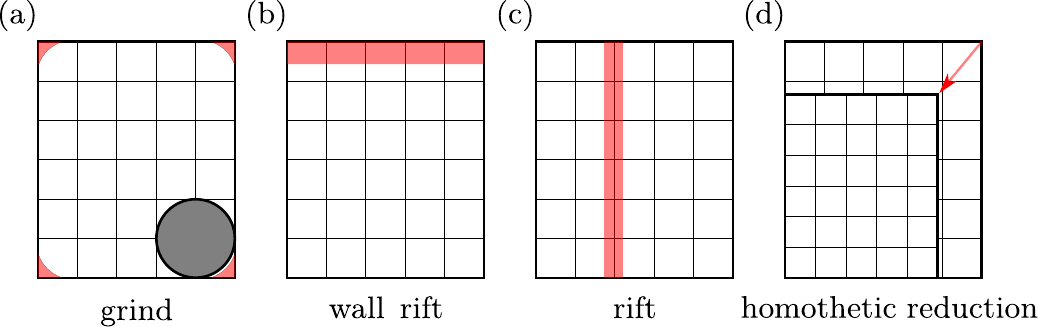}
\caption{Volume reductions for a \np  box.
\subcap{a} Pathological corner-grind volume reduction, that
eliminates no samples for 
sufficiently large $\sigma$.
\subcap{b} Horizontal wall rift used to estimate the 
pressure $P_y$.
\subcap{c} Vertical rift used to estimate $P_x$.
\subcap{d} Homothetic volume reduction. }
\label{fig:VolumeChanges} \end{figure}

\subsubsection{Equation of state, phase coexistence}
\label{subsec:EquationOfState}

In the thermodynamic limit, the stability of  matter is expressed through a 
positive compressibility $\kappa 
= -(\partial V / \partial P) / V$. For a finite system in the $NVT$ 
ensemble, this is not generally true in a presence of a first-order phase 
transition, where two coexisting phases are separated by an 
interface with its own interface free energy.

\begin{figure}[htb]
\centering
\includegraphics[width=\columnwidth]{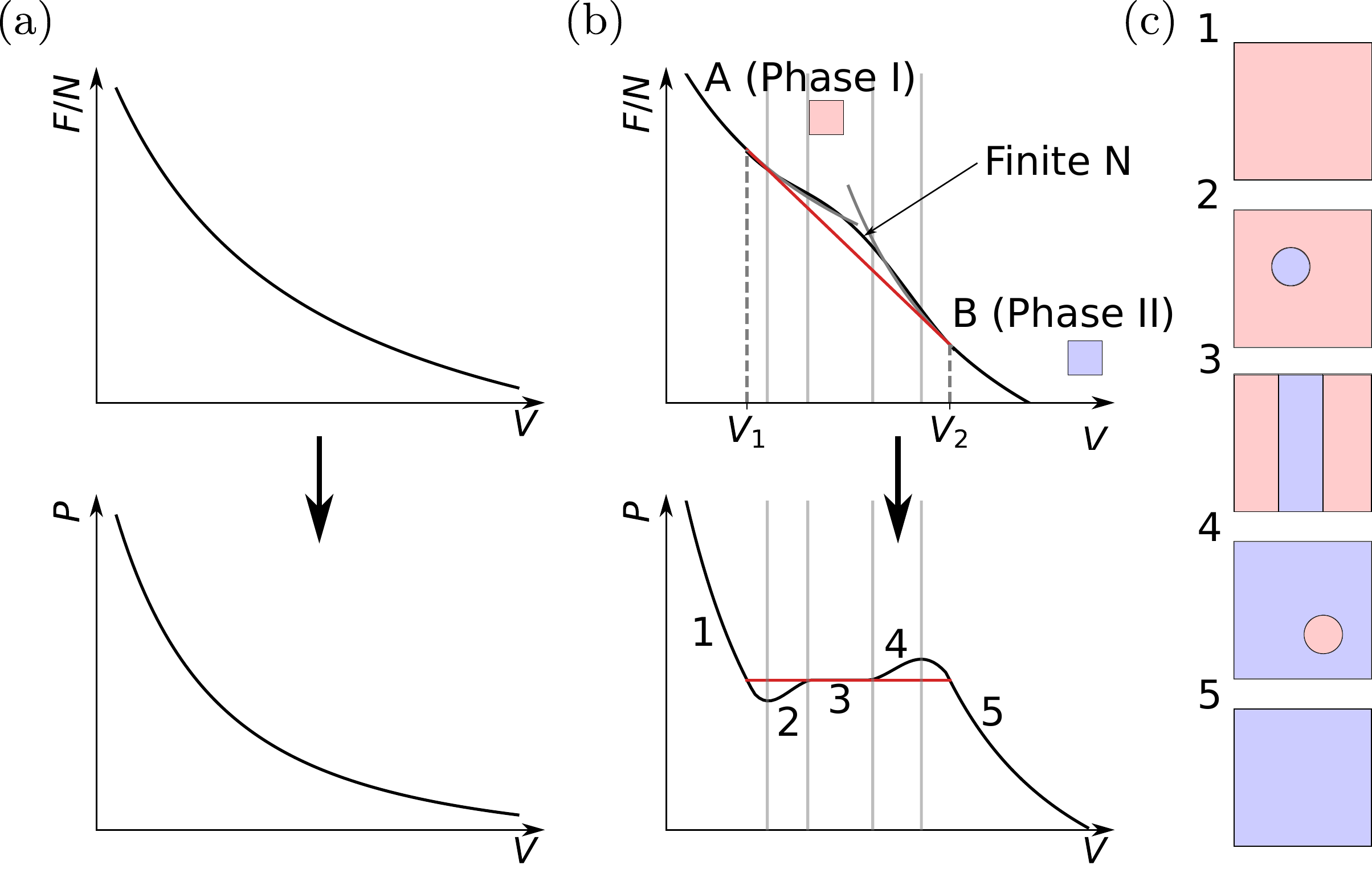}
\caption{First-order phase transition in the $NVT$ ensemble. \subcap{a}  Free 
energy with increasing second derivative, and thus a monotonously decreasing 
pressure. 
\subcap{b}  Free energy with---for the infinite system---metastable branches 
starting at volumes $V_1$ and $V_2$, and a non-monotonous equilibrium pressure 
$P$ for finite $N$. \subcap{c} Sequence of five regimes in a finite 
two-dimensional periodic box, with pure (1,5), bubble (2,4) and stripe (3)
phases. 
}
\label{fig:Phases}
\end{figure}

In a periodic two-dimensional  box for finite $N$, on increasing the density 
(decreasing the volume), a first-order phase 
transition first creates a bubble of the denser phase in the less dense phase 
(for hard disks: a hexatic bubble inside the fluid). The stabilization of this 
bubble requires an extra  \quot{Laplace} pressure corresponding to the surface 
tension, which renders the overall pressure non-monotonic with 
$V$~\cite{MayerWood1965}. At
larger densities, 
the bubble of the minority phase transforms into a stripe that winds 
around the periodic box. In the stripe regime, the length of the two interfaces 
and thus the interface free energy do not change with $V$, so that the pressure 
remains approximately constant. Finally, one obtains a bubble of the less dense 
phase (the fluid) in the surrounding denser (hexatic) phase (see 
\fig{fig:Phases}). 
The 
phase coexistence is specific to the $NVT$ 
ensemble as certain specific volumes $V/V_0$ do not correspond to densities 
$\eta  = (N/V) \pi \sigma^2$ of a homogeneous stable phase for $N \to \infty$. 
Phase coexistence is absent in the $NPT$ ensemble. The pressure is then the 
control variable, and the volume $V / V_0$ is discontinuous at the transition,  
providing for a simpler physical picture. However, in the $NPT$ ensemble and its 
variants, sampling algorithms generally converge even more slowly than in the 
$NVT$ ensemble. 

The phase coexistence and the non-monotonous equation of state are genuine 
equilibrium features at finite $N$. Moreover, the spatially inhomogeneous 
phase-separated equilibrium state is reached from homogeneous initial 
configurations through a slow coarsening process, whose dynamics
depends on the sampling algorithm.

\section{Sampling algorithms and pressure estimators}
\label{sec:Algorithms}

In the present section, we discuss sampling algorithms (molecular dynamics, 
MCMC) and pressure estimators for the hard-disk model. All 
algorithms are unbiased and correct to machine precision. For example, molecular 
dynamics implements the Newtonian time evolution of disks without discretizing 
time. The reversible Monte Carlo algorithms, from the historic Metropolis 
algorithm to the recent massively parallel Monte Carlo (MPMC) on graphics-card 
computers, satisfy the detailed-balance condition: the net equilibrium flow 
vanishes between any two samples. The non-reversible ECMC algorithms only 
satisfy the global-balance condition, and  samples live in an 
extended \quot{lifted} sample space.

Besides the sampling algorithms, we also discuss pressure estimators, including 
recent ones that overcome the limitations of the traditional approaches. 
Thermodynamic pressure estimators compute the probability with which a sample is 
eliminated as the box is reduced in size while kinematic pressure estimators 
determine the momentum fluxes at the walls or inside the box. Importantly, 
both types of estimators have the same expectation value (the pressure $P$ or 
its components $P_x$ or $P_y$), and they merely differ in their efficiency and 
ease of use. Even at finite $N$, there is thus no ambiguity in the definition of 
the pressure. The various estimators play a key role in the equation-of-state 
computations in \sect{sec:EquationsOfState}. All algorithms and 
estimators are cross-validated for four disks in a 
\np square box (see \app{subsec:FourDisksInSquare}).
These tiny simulations illustrate
the absence of any finite-$N$ bias in the pressure estimators.

\subsection{Event-driven molecular dynamics (EDMD)}
\label{subsec:MolecularDynamics}

Event-driven molecular dynamics (EDMD)~\cite{Alder1957} implements the 
Newtonian time evolution for the hard-disk model by stepping forward from one 
event (pair collision or wall collision) to the next.
Between collisions, the disks move
with constant velocities. Collisions of more than two disks, or 
simultaneous collisions can be neglected. For large run times $\tsim$, 
if the sample space is connected,
molecular dynamics samples the equilibrium 
distribution of positions and velocities of \eq{equ:hypersphere}. 

\subsubsection{Naive molecular-dynamics program}
\label{subsec:NaiveMD}
From a given set of positions and velocities for $N=4$ disks in a \np 
box, our 
naive EDMD code  computes the minimum over all pair collision times for the 
$N(N-1)/2$ pairs of disks, and over the wall collision times for the $N$ disks. 
This 
minimum corresponds to a unique collision event (multiple overlaps appearing 
with finite-precision arithmetic can be treated in an ad-hoc fashion). The code 
then 
updates all the positions and the velocities of the colliding disks. This 
algorithm is of complexity \bigOb{N^2} per event. A related naive program in a 
periodic box (with arbitrary $N$) is used for cross-validation of other 
algorithms. Practical implementations of  EDMD process the collisions through 
floating-point arithmetic. As the hard-disk dynamics is chaotic for almost all 
initial configurations, trajectories for different precision levels quickly 
diverge 
and only the statistical properties of the trajectories are believed to be  
correct. 

\subsubsection{Modern hard-disk molecular dynamics}
\label{subsec:ModernEDMD}

The complexity of EDMD can be reduced from the naive \bigOb{N^2} per 
event scaling to 
\bigOb{\log N}. This is because the collisions of a given disk must 
only be tracked with other disks in a local neighborhood (reducing by itself 
the complexity to \bigOb{N}) and because the collision of two disks $i$ and $j$ 
only modifies the future collision times for pairs involving $i$ or $j$ 
(see~\cite{Rapaport1980,Rapaport2009}). This algorithm keeps \bigOb{N} candidate 
events of which a finite number must be updated after each event. Using a
heap data structure, this is of complexity \bigOb{\loga{N}}, while the retrieval 
of the shortest candidate event time (the  next event time) is \bigOb{1}. 
Although the update of the event times involves elaborate book-keeping, and 
although the processing of events according to collision rules is 
time-consuming, the EDMD 
algorithm is thus fairly efficient. It has been successfully used 
for the hard-disk model up to intermediate sizes ($N\lesssim 256^2$ in the 
transition region). Open-source implementations of this algorithm are 
available~\cite{Bannerman2011}. 
The EDMD algorithm has not been successfully parallelized,
despite some efforts in that direction~\cite{lubachevsky1992simulating}.

\subsection{Hard-disk Markov-chain Monte Carlo}
\label{subsec:MonteCarlo}

Hard-disk Monte-Carlo algorithms consider a sample space consisting of the $N$ 
positions. Initial samples that are easy to construct,  are 
modified through 
reversible or non-reversible schemes. In the large-time limit,  the 
equal-probability measure of the positions in \eq{equ:hypersphere} is reached. 

\subsubsection{Local hard-disk Metropolis algorithm}
\label{subsec:Metropolis}

In the local hard-disk Metropolis algorithm~\cite{Metropolis1953}, at each time 
step, a small random displacement of a randomly chosen disk is accepted if the 
resulting sample is legal and is rejected otherwise. A move and its 
inverse are proposed with the same probability, so that the algorithm satisfies 
the detailed-balance condition with a constant equilibrium probability, and the 
net probability flows vanish. 

The local Metropolis algorithm has been much used to obtain the 
hard-disk equation of state. On a modern single-core central-processing unit 
(CPU), this algorithm realizes  
roughly $\sim 10^{10}$ moves per hour. (For simplicity, we use \quot{moves}
for \quot{proposed moves}.)
However, its convergence is very slow. In 
\sect{subsec:LargeSystems}, 
we will show evidence of 
mixing times~\cite{Levin2008} in excess of 10 years of CPU time for $\sim 10^6$ 
disks.
The 
sequential variant of the local Metropolis algorithm updates the disk $i+1$ 
(identifying $N+1 \equiv 1$) at time $t+1$ after having updated disk $i$ at time 
$t$. This non-reversible version runs slightly faster as it requires fewer 
random numbers per move, but the performance gain is minimal. 

\subsubsection{Massively parallel Monte Carlo (MPMC) algorithm}

\begin{figure}[htb]
\centering
\includegraphics[width=0.5\columnwidth]{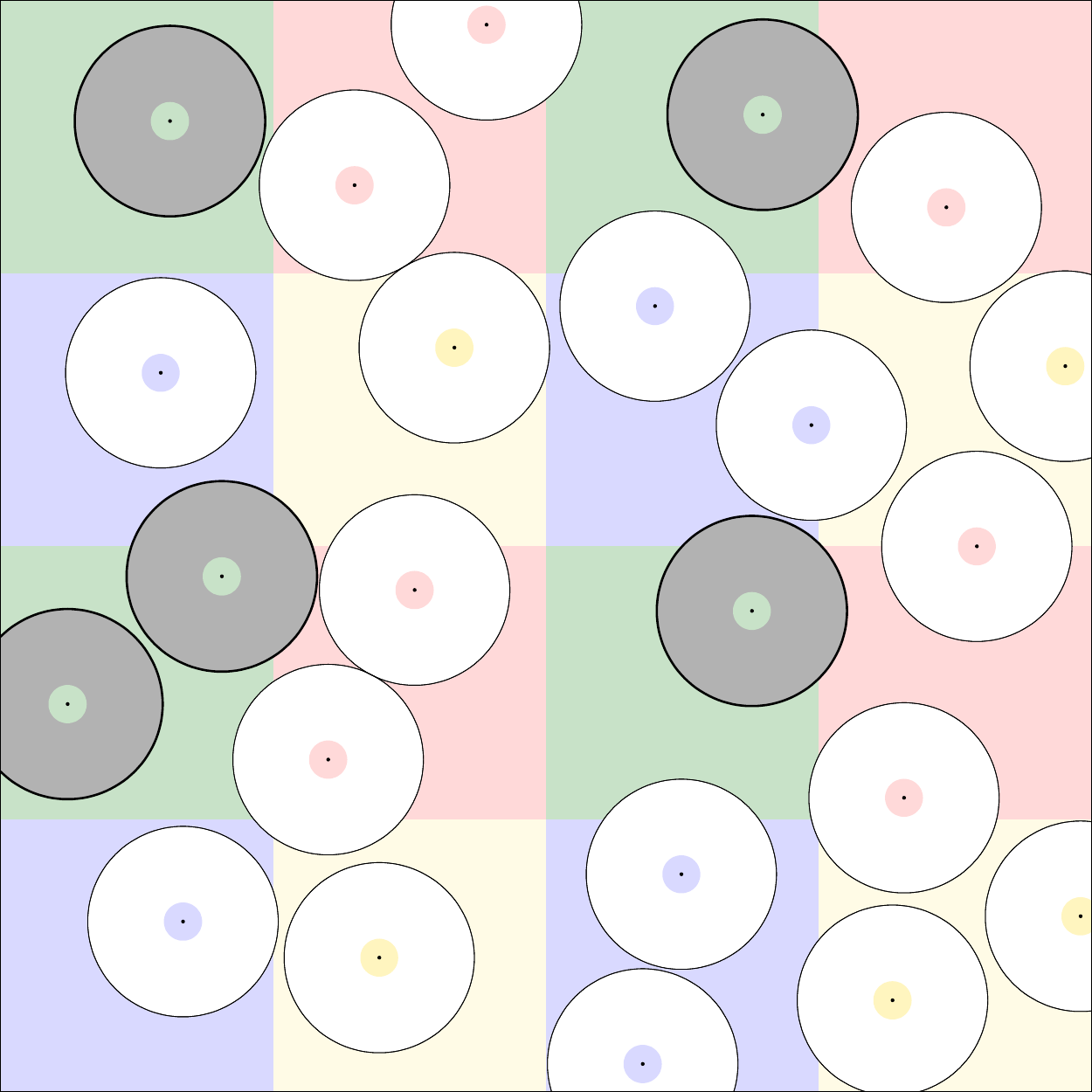}
\caption{Four-color checkerboard decomposition in a periodic box, with cells  
larger than $2\sigma$. If the \emph{green} color is chosen, highlighted disks 
may move, but cannot leave their cells. Disks in different green cells do not 
communicate.}
\label{fig:ParaMetSchema}
\end{figure}

The MPMC algorithm generalizes the local 
Metropolis algorithm for implementation on  graphical processing units 
(GPU)~\cite{AndersonGPU2013}. It uses a four-color checkerboard of rectangular 
cells of sides larger than $2 \sigma$, that is superposed onto the periodic 
box and is compatible with the periodic boundary conditions.
Cells of the same color are distant by more than $2 \sigma$. They are
aligned with the $\ehatvec_x$ and $\ehatvec_y$ axes.
The MPMC algorithm samples one of the 
four colors, and then independently updates  disks in all corresponding cells 
using the local Metropolis algorithm with the additional constraint that disks 
cannot leave their cells (see \fig{fig:ParaMetSchema}). After a certain time, 
the color is resampled. The checkerboard is frequently detached from the 
box, then randomly translated and repositioned, rendering the algorithm 
irreducible.

On a single NVIDIA GeForce RTX3090 GPU, our MPMC code reaches $2.1 \times 
10^{13}$ moves per hour, an order of magnitude more than an 
earlier implementation~\cite[Table II]{Engel2013}. Repositioning the 
checkerboard is computationally  cheap and is done often enough 
for the convergence time, measured in moves, to be only 
slightly 
larger than for the local Metropolis algorithm. 

\subsubsection{Hard-disk event-chain Monte Carlo (ECMC)}

Hard-disk ECMC is a non-reversible continuous-time \quot{lifted} Markov 
chain~\cite{Bernard2009} in which---on a single processor---at each time $t$, a 
single active disk moves with constant velocity, while all the others are at 
rest. The identity and velocity of the active disk constitute 
additional \quot{lifting variables} in an extended (lifted) sample 
space~\cite{Chen1999,Diaconis2000,Krauth2021eventchain}. 
At a collision event, the active disk stops, and the target disk 
becomes active. The  variants of ECMC differ in how the velocity is 
updated. 
In the straight variant, the velocity of the active disks is maintained after a 
pair-collision event. It is usually 
chosen to be either in the $ \pm \ehatvec_x $ or the $\pm \ehatvec_y$ 
direction, that 
is, along one of the coordinate axes. At a wall-collision event, the 
velocity 
is flipped, for example from $\pm \ehatvec_x$ to $\mp \ehatvec_x$.
In addition, 
resampling events  take place typically at equally spaced times 
separated by the run-time interval $\tsim$.
At such resamplings, 
a new active disk is sampled, and the new velocity is sampled from 
$ \SET{\pm \ehatvec_x, \pm \ehatvec_y}$.
With periodic boundary conditions, the new velocity is sampled from 
$ \SET{+\ehatvec_x,  +\ehatvec_y}$.

Under conditions of irreducibility and aperiodicity, ECMC samples the 
equilibrium distribution of hard-disk positions with non-zero net probability 
flows. 
However, the hard-sphere ECMC and the hard-sphere 
local Metropolis algorithm are not strictly irreducible~\cite{Hoellmer2022}.
ECMC is much more powerful than the local Metropolis algorithm, and in 
\sect{subsec:LargeSystems}, 
we will evidence speedup factors of $\sim 10^3$.
The \HD software package (see 
\app{app:DataCodes}) contains sample codes for straight ECMC,  reflective 
ECMC~\cite{Bernard2009}, forward ECMC~\cite{Michel2020}, and Newtonian 
ECMC~\cite{Klement2019}. Straight ECMC is fastest for the 
hard-disk model, and it was successfully parallelized~\cite{Li2020}. 
The performance of straight ECMC is roughly of $10^{10}$ events (collisions) 
per hour on a modern single-core CPU. Its parallelized version reaches 
$\lesssim 10^{11}$ events per hour. This performance is currently limited by a 
hardware bandwidth bottleneck~\cite{LiNishikawaMaggsKrauth2022}, that will be 
overcome in the near future.

\subsection{Hard-disk pressure estimators}
\label{subsec:PressureComputations}
As discussed in \subsect{subsubsec:PressurePartitionFunction}, the pressure
describes, on the one hand,  the change of the free energy when the
volume is reduced and, on the other hand, the time-averaged momentum exchange 
with the walls. In the present subsection, we reduce the volume 
through rifts and rift averages, and by uniformly shrinking the box. 
We also compute the momentum exchange directly and through a virial formula. 
Our motivation is two-fold. First, we obtain practical pressure 
estimators that we implement in our algorithms. 
Second, we discuss in detail that all the pressure estimators 
of Monte Carlo and of molecular dynamics compute the same object, and 
this even for finite systems. The decades-long discrepancies in the estimated 
pressures can thus not be traced to differences in their definitions.

\subsubsection{Rifts and rift averages}
\label{subsec:Chipping}

\begin{figure}[htb]
\centering
\includegraphics[width=0.7\columnwidth]{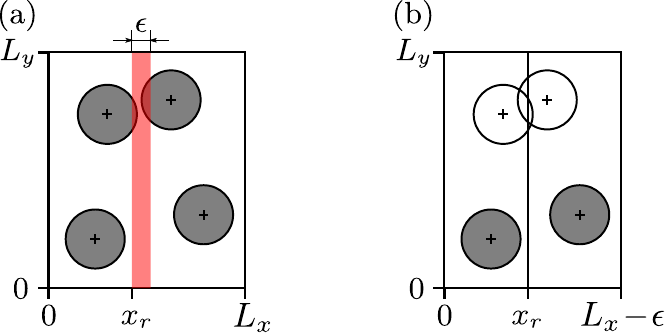}
\caption{Vertical rift $[x_r,x_r+\epsilon] \times [0,  L_y]$. \subcap{a} $L_x
\times
L_y$ box with vertical rift of width $\epsilon$ at position $x_r$. \subcap{b}
Transformed sample, which is eliminated because of a pair overlap.}
\label{fig:Rift}
\end{figure}

In an $L_x \times L_y$  box, the volume may 
be reduced through a vertical
\quot{rift} $[x_r, x_r+ \epsilon]\times [0,L_y]$
with disk positions transforming as:
\begin{equation}
\SET{x, y} \to \begin{cases}
            \SET{x,y} &  \text{if $x< x_r$}\\
            \emptyset &  \text{if $x_r \le x< x_r + \epsilon$}\\
            \SET{x-\epsilon,y} &  \text{if $x \ge x_r+ \epsilon$},
           \end{cases}
\label{equ:RiftDefinition}
\end{equation} 
where \quot{$\emptyset$} means that the position is eliminated.
A rift either transforms a uniform hard-disk sample into a uniform sample in 
the 
reduced box, or else eliminates it because a disk falls inside the rift 
or because two disks overlap (see \fig{fig:Rift})
In a \np box, wall rifts
at $x_r =0$ or $x_r = L_x - \epsilon$  (and likewise for $y$) chip off a slice 
from the surface.
Vertical rifts, 
as in \eq{equ:RiftDefinition}, estimate the pressure $P_x$, and horizontal 
rifts ($[0, L_x] \times [y_r, y_r+ \epsilon]$) the pressure $P_y$. 
Simultaneous vertical and horizontal rifts
with $L_y \epsilon_x = L_x \epsilon_y$
conserve the aspect ratio of the  box.  Equivalent to a homogeneous
(homothetic) rescaling of the box, they estimate the pressure $P = (P_x
+ P_y)/2$.
The pressure can be estimated for finite $\epsilon$ from a finite number of 
samples, but then requires an extrapolation towards  $\epsilon \to 0$. In EDMD 
and ECMC, the extrapolation can be avoided because of the infinite number of  
samples produced in a given run-time interval $\tsim$.

We first reduce the volume $V$ of a \np $L_x \times L_y$  box by
a vertical wall rift with $x_r = L_x - \epsilon$ or
by a horizontal wall rift with $y_r = L_y - \epsilon$.
A sample in the original box
is eliminated through the wall rift
with the probability that one of the disks
overlaps the wall rift.
Disk $i$ is at position $\xvec_i$
with the normalized single-disk probability 
density
$\piOne(\xvec_i)$ and at an $x$-position 
in the interval  $[L_x - \sigma - \epsilon,
L_x - \sigma]$ with probability
$\epsilon \int \diff y 
\piOne(L_x - \sigma,y)$. 
We normalize the single-disk density $\rho$ to 
$V$, so that the normalized probability density $\piOne(\xvec_i)$ of a given 
disk 
$i$
to be at position 
$\xvec_i$ equals $\pi(\xvec_i) = \rho(\xvec_i)/V$.
We then use the rescaled line density $\rhox(x) = 
\int \diff y \rho(x, y) /L_y$, and likewise for $\rhoy(y)$. 
With the vertical rift volume $\epsilon L_y$ and analogously for the 
horizontal one, this gives the wall-rift pressure estimator:
\begin{equation}
\beta \glc
\begin{array}{l} P_x\\
P_y
\end{array}
\grc = \frac{ N}{V} \glc 
\begin{array}{l}
\rhox(L_x - \sigma) \\
\rhoy(L_y - \sigma) 
\end{array}
\grc.
\label{equ:PressureEarFormula}
\end{equation}
Naively, the rescaled line densities $\rhox$  and $\rhoy$
are obtained
from a histogram of the $x$-coordinates, extrapolated to $x = L_x - 
\sigma$ and equivalently, to $x= \sigma$ (see \tab{tab:UltraNaive}, line 1 
and \app{app:Statistics}). 

\newcommand{\eqsee}[1]{(see \eq{#1})}
\begin{table}[ht]
\centering
\begin{tabular}{lll}
\hline
\#& Sampling method: pressure estimator & $\beta P$ \\
[0.5ex]
\hline
1& EDMD: wall-rift fit \eqsee{equ:PressureEarFormula} & $10.74(7)$ \\
2& EDMD: wall rift \eqsee{equ:PressureMDWallc} & $10.79625(4)$ \\
3& ECMC: wall rift \eqsee{equ:PressureEarECMC} & $10.7962(4)$ \\
4& EDMD: rift average \eqsee{equ:MDswellCollisiona} & $10.79629(3)$ \\
5& ECMC: rift average \eqsee{equ:ff-w} & $10.7962(4)$ \\
6& EDMD: homothetic fit \eqsee{equ:PressureSwellFiniteA} & $10.74(4)$ \\
\hline
\end{tabular}
\caption{Thermodynamic pressure estimates for four disks of radius 
$\sigma=0.15$ in a 
\np square box of sides $1$.
The kinematic estimators of \subsect{subsec:KinematicEstimators} lead to 
identical expressions. 
} 
\label{tab:UltraNaive}
\end{table}

Within EDMD, the rescaled line densities of 
\eq{equ:PressureEarFormula} 
can be computed, without extrapolation, from the time 
interval $ \Delta t = 2 \epsilon / |\vperpwall|$ before and after the 
collision during which a disk with perpendicular velocity $\vperpwall$ overlaps 
with 
the wall rift at $x_r = L_x -  \epsilon$.
The time interval $\Delta t$ here simply 
indexes equilibrium samples and has no kinematic meaning.
The change of volume (by the two rifts 
at $x=0$ and $x  = L_x$) equals $2 \epsilon L_y$. For $\epsilon \to 0$, 
only a single disk overlaps with the wall rift, leading to the EDMD wall-rift 
estimator:
\begin{subequations}
\begin{align}
\beta P_x  & = 
\frac{1}{2L_y\tsim}\sum_{\wallcolls[,\pm \ehatvec_x]{i}} \frac{
2}{|\vperp_{(i)}|}\quad \label{equ:PressureMDWalla}\\
& = 
\mean{\frac{
2}{|\vperp_{\text{wall}}|} }
\overbrace{
\frac{1}{2L_y\tsim}
\sum_{\wallcolls[,\pm \ehatvec_x]{i}} 1}^{\nunitwallx}
\label{equ:PressureMDWallb}\\
& = \frac{2\sqrt{\pi}}{\sqrt{\sum v_i^2}}\frac{\Gamma(N + \half)} {\Gamma(N)}
\nunitwallx
\label{equ:PressureMDWallc} \\
 & \simasymp \sqrt{2 \pi \beta m} \nunitwallx.
\label{equ:PressureMDWalld}
\end{align}
\label{equ:PressureMDWall}%
\end{subequations}
The sum in \eq{equ:PressureMDWalla} goes over the wall collisions $w$ of all 
disks $i$ in $\pm \ehatvec_x$ direction, and $\nunitwallx$ is the 
wall-collision rate per vertical unit line element.
In \eq{equ:PressureMDWall}, the right-hand sides are estimators, whose 
expectation 
yields the pressure on the left-hand side. In this equation and the following 
ones, the additional
$\mean{\dots}$ (such as $\nunitwallx \to 
\mean{\nunitwallx}$ in \eqtwo{equ:PressureMDWallc}{equ:PressureMDWalld}) are 
omitted.
In 
\eq{equ:PressureMDWalla}, the estimator has infinite variance. It is 
regularized through its mean value
in \eq{equ:PressureMDWallb}. The latter is evaluated in 
\eq{equ:PressureMDWallc} with the Maxwell-boundary expression of 
\eq{equ:MaxwellBoundary}, and tested to $\fpn{4}{-6}$ in relative precision 
against other estimators (see \tab{tab:UltraNaive}, line 2).
The pressure 
estimator of \eq{equ:PressureMDWallc} 
can also be derived as a 
kinematic pressure estimator through the momentum transfer with the walls (see 
\subsect{subsec:KinematicEstimators}). Thermodynamic and kinematic pressures 
thus agree already at finite $N$.

\begin{figure}
\includegraphics[width=0.95\columnwidth]{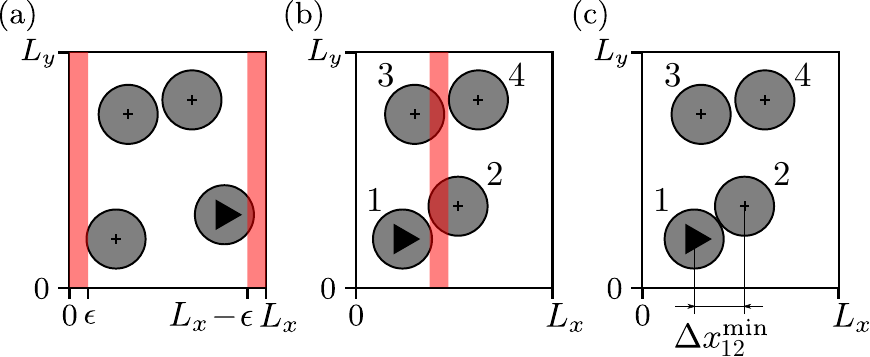}
\caption{ECMC rift estimators.
\subcap{a} The ECMC wall-rift estimator only detects rift overlaps of the 
active disk, explaining the factor $N$ in \eq{equ:PressureEarECMC}, that is 
absent in \eq{equ:PressureMDWalla}.
\subcap{b} A pair of disks $(i,j)$ leading to the elimination of the sample is 
detected only if either $i$ or $j$ are active, explaining a factor $N/2$
entering \eq{equ:ff-w}. 
\subcap{c} Illustration of the $x$-separation at contact $\delxminij$ (also 
relevant for EDMD).
}
\label{fig:BiasECMCMD.png}
\end{figure}

The wall-rift pressure estimator adapts non-trivially 
to ECMC. We consider straight ECMC with a single active disk 
that  moves with 
unit speed along the $\pm \ehatvec_x$ direction. As a lifted 
Markov chain, 
ECMC splits the equilibrium probability of each \quot{collapsed} 
sample $\xvec$ equally between the $N$ lifted 
copies (consisting of $\xvec$ and of the label of the active 
disk for a given displacement vector) (see~\cite{Krauth2021eventchain} 
and~\cite[Appendix A]{Qin2022}). ECMC only 
determines overlaps with the walls for the active disk, 
and a lifted sample that must be eliminated is 
detected 
with a biased probability $1/N$. This bias is corrected by multiplying the
right-hand side of \eq{equ:PressureMDWalla} by $N$, resulting in the ECMC 
wall-rift estimator: 
\begin{equation}
    \beta P_x = \frac{N}{2 L_y\tsim}\sum_{\wallcolls[,\pm \ehatvec_x]{a}}
    \frac{2}{|\vperpwall|} 
    = 2N \nunitwallx 
\label{equ:PressureEarECMC}%
\end{equation}
(see \subfig{fig:BiasECMCMD.png}{a}).
It is tested to $\fpn{4}{-5}$ in relative precision (see \tab{tab:UltraNaive}, 
line 3).

Within molecular dynamics and ECMC, pressures can also be estimated by rifts 
inside the box and in particular by averages over all rift positions $x_r$  
in addition to the average over $\tsim$ already contained in the wall-rift 
estimators. This can be written as 
\begin{equation}
\beta P_x  =
\frac{1}{\epsilon L_y}
 \frac{1}{L_x \tsim}  \int_{t}^{t + \tsim } \diff \tau \int_0^{L_x - \epsilon}
\diff x_r \Theta(\tau,x_r),
\label{equ:ECMCBoxPressure}
\end{equation}
where $\Theta(\tau,x_r)$ is zero if the sample at time $t + \tau$ is 
maintained after the reduction with parameter $x_r$ and one if it is 
eliminated. The ideal-gas contribution to \eq{equ:ECMCBoxPressure},
\begin{equation}
\beta P_x^{\text{ideal gas}} =
\frac{\epsilon N}{L_x} \frac{1}{\epsilon L_y} =
\frac{N}{V},
\end{equation}
counts the proportion of rifts that eliminate samples because 
the centers of the disks fall inside. 
The pair-collision contribution to \eq{equ:ECMCBoxPressure} is derived 
considering the sample shown in \fig{fig:Rift}.
If the distance between two disks $i$ and $j$ is in the interval $[2 \sigma, 
2 \sigma + 
\epsilon \delxminij /(2 \sigma)]$, where $\delxminij$ is the 
$x$-separation at contact,
the corresponding samples are eliminated for a time 
$(2 /|  \vrelperppair |)[\epsilon \delxminij/(2 \sigma)]$
for vertical rifts in the interval of length 
$\delxminij$ between the two disks at contact.
Together  with the wall term analogous to \eq{equ:PressureMDWall}, the 
rift-average 
pressure estimator for EDMD thus reads: 
\begin{align}
\beta P_x = \frac{N}{V} + \frac{1}{V \tsim} & \glc \sum_{\paircolls} 
\frac{|\delxminij|^2 
}{2 \sigma}\right. \mean{\frac{2}{ \vrelperppair}} 
\nonumber\\
&\left. +\sum_{\wallcolls[,\pm \ehatvec_x ]{i}} 
\mean{
\frac{2\sigma}{|\vperp_{\text{wall}}|}} \grc, 
\label{equ:MDrift}
\end{align}
where the mean values  again involve Maxwell-boundary 
expressions. 
\Eq{equ:MDrift} can be 
combined with an analogous expression for $P_y$ to obtain
the EDMD rift-average estimator for $P$: 
\begin{align}
\beta P = \frac{N}{V} + \frac{\sigma}{V \tsim}\glc \sum_{\paircolls} 
\right. &\mean{\frac{2}{ \vrelperppair}}
\nonumber\\
&\left. +\sum_{\wallcolls[,\pm \ehatvec_x, \pm \ehatvec_y]{i}} 
\mean{ \frac{1}{|\vperp_{\text{wall}}|}}\grc. 
\label{equ:FixLabelWK}
\end{align}
In a \np box, using \eqtwo{equ:MaxwellBoundary}{equ:MaxwellPair}, the 
EDMD rift-average estimator takes the form
\begin{subequations}
\begin{align}
\beta P &= \frac{N}{V} + \frac{\sigma \sqrt{\pi}}{\sqrt{\sum v_i^2}V}
\frac{\Gamma(N + \half)}{\Gamma(N)}
(\nwallxy + \sqrt{2} \npair) 
 \label{equ:MDswellCollisiona} \\
   & \simasymp
 \frac{N}{V} \glb 1 + \frac{\sigma \sqrt{\pi m \beta}}{N} \npair \grb,
 \label{equ:MDswellCollisionb}
\end{align}
\label{equ:MDswellCollision}%
\end{subequations}
where $\nwallxy$ is the wall-collision rate, the number of all wall collisions 
per time interval, and similarly for the pair-collision rate $\npair$.
In the $N \to \infty$ limit of 
\eq{equ:MDswellCollisionb}, wall collision play no role.
The EDMD  rift-average estimator of
\eq{equ:MDswellCollisiona} is tested to $\fpn{3}{-6}$ in relative precision 
(see \tab{tab:UltraNaive}, line 4).

Rift-average pressure estimators for  ECMC 
detect wall and pair collisions with biases (see \eq{equ:PressureEarECMC}), 
that must again be corrected, namely  by  a factor $N$ for each wall event and 
by a factor $N/2$ for each pair event, the latter because
a lifted sample of $N$ disks that must be eliminated 
is detected only if either $i$ or $j$ are active (see
\subfig{fig:BiasECMCMD.png}{c}). This leads to
the straight-ECMC rift-average estimator, 
\begin{equation}
\beta P_x = \frac{N}{V} + \frac{N}{V \tsim}\glb  
\sum_{\paircolls}\delxminij +
\sum_{\wallcolls[\pm \ehatvec_x]{i}} 2\sigma \grb, 
\label{equ:ff-w}
\end{equation}
that again differs in the factors $\propto N$ from the corresponding formulas 
of EDMD. Furthermore, it averages over a bounded distribution of 
$\delxminij$, with the wall-velocity only taking the values $\pm 
1$, 
whereas in EDMD, the corresponding continuous distributions of $1/| 
\vrelperppair| $ and 
of $1/|\vperpwall|$ have infinite variance. The straight-ECMC 
estimator of 
\eq{equ:ff-w} is tested to $\fpn{4}{-5}$ in relative precision 
(see \tab{tab:UltraNaive}, line 5).

In a periodic box,  
there are no wall collisions,
and the direction of motion of straight ECMC
is either $+\ehatvec_x$ or  $+\ehatvec_y$.
In \eq{equ:ff-w}, all
the $x$-separations at contact $\delxminij$ and the 
chain length $\tsim$ (which, because of the unit velocity, corresponds to the 
total displacement) add up to the difference of the final position 
$x_\text{final}$ of the last disk of the 
chain and the initial position $x_\text{initial}$ of the chain's first disk. 
Here,  
periodic boundary conditions are accounted for, so that
in the absence of collision, this distance equals $\tsim$.
For an event chain in the $+\ehatvec_x$ direction, \eq{equ:ff-w} thus 
simplifies into the straight-ECMC estimator 
for a periodic box~\cite{Michel2014JCP}:
\begin{equation}
\beta P_x = \frac{N}{V \tsim}(x_{\text{final}} -
x_{\text{initial}}), 
\label{equ:RiftPressure}
\end{equation}
that is  easy to compute, and that will be used extensively in 
\sect{sec:EquationsOfState}. There, we alternate
event chains in $+ \ehatvec_x$, which
estimate $P_x$, and
event chains in $+\ehatvec_y$, which estimate
$P_y$. 
Alternating the 
direction of straight-ECMC chains is required for convergence towards 
equilibrium.
The rift-average estimator generalizes to other variants of ECMC.
The pressure $P_y$ can also be estimated through  
event chains in $\pm 
\ehatvec_x$
and horizontal-rift averages, leading for 
the straight ECMC in $+\ehatvec_x$ to:
\begin{equation}
 \beta P_y =  \frac{N}{V} + \frac{N}{V \tsim} \sum_{\paircolls} \frac{|\delymin
_{ij}|^2}{\delxminij}.
\end{equation}
However, this estimator has infinite variance and is less convenient than 
\eq{equ:RiftPressure}.

\subsubsection{Homothetic volume reductions}
\label{subsec:ShrinkingPair}

Besides by rifts, the volume $V$ of an $L_x \times L_y$ box
can be reduced by a homothetic transformation, where the box
and all positions $\xvec_i$ are homogeneously scaled by a
factor $1 - \epsswell< 1$, while the disk radii $\sigma$ remain unchanged. 
The transformation of the box corresponds to 
simultaneous
horizontal and vertical rifts of equal rift volume, but the disk positions then
transform inhomogeneously, as in \eq{equ:RiftDefinition}. 

\begin{figure}[htb]
\centering
\includegraphics[width=0.7\columnwidth]{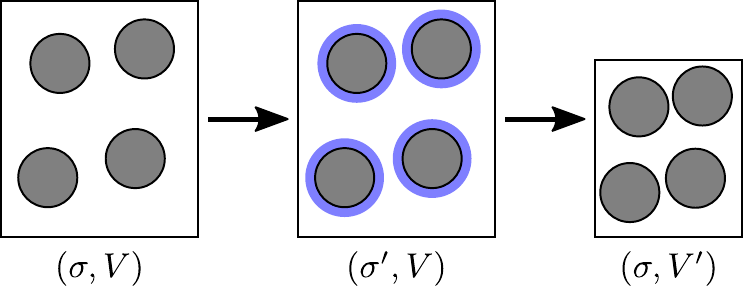}
\caption{A homothetic volume reduction performed through a swelling of disks
followed by a uniform reduction of all lengths (box, positions, radii).
}
\label{fig:ShrinkBox}
\end{figure}

A homothetic volume reduction yields  the pressure $\beta P =
\beta( P_x + P_y)/2$, rather than one of the components.
It may be performed in two steps. In a first step 
(from $(\sigma, V)$  to $(\sigma',V)$, see \fig{fig:ShrinkBox}), the box and 
the $\xvec_i$ are unchanged, but the disks are swollen by a 
factor $1/(1 - \epsswell)$, possibly eliminating samples.
In a 
second step, all lengths are rescaled by $1 - \epsswell$, 
so that the radii return to 
$\sigma$. This second step (from  $(\sigma', V)$  to $(\sigma, V')$)  
is rejection-free, and its reduction of sample-space volume, with $ Z 
(\sigma,V') 
= 
(V'/V)^N Z(\sigma', V)$, constitutes the ideal-gas term of the 
pressure.

In the two-stage transition
$Z(\sigma, V) \to Z(\sigma', V) \to Z(\sigma, V')$,
the two-step procedure turns \eq{equ:PressureElimination} into
\begin{equation}
\begin{aligned}
\beta P \stackrel{V'\to V}{=}&  \frac{\logc{Z(\sigma, V)} - \logc{Z(\sigma, 
V')}}{V - V'} \\
=& \frac{N}{V} + \frac{1}{ V - V'} \frac{Z(\sigma, V) - Z(\sigma', V)
}{Z(\sigma, V)}.
\label{equ:pressureswell}
\end{aligned}
\end{equation}
The final term again divides the elimination probability of a sample by the 
change of volume (see also~\cite{Eppenga1984,Brumby2011,Miguel2006,Allen2006}).

The pair-elimination probability is expressed by the normalized 
probability density
$\ghat(r_x,r_y)$, with $\ghat(r_x,r_y) \diff r_x \diff r_y$ the probability 
that a that a given pair distance is in $[r_x,r_x + \diff x][r_y,r_y+ \diff 
y]$. With $\ghat(r)$ as the average of $ \ghat(r_x,r_y)$ over the corresponding 
ring of radius $r$, 
the probability that a given pair distance
is in the interval $[r, r+ \diff r]$ is 
thus $2\pi r \ghat(r) \diff r$. By convention, the pair correlation function 
$g(r)$
is normalized to $g(r) = V \ghat(r)$. For our 
application, we have $r = 2 \sigma$ and  $\diff r = 2 \sigma \epsswell$, and 
there are $N(N-1)/2$ pairs of disks. Also, the absolute change of volume for 
$L_x \to L_x( 1 + \epsswell)$ and  $L_y \to L_y( 1 + \epsswell)$ is $2 V 
\epsswell$. The pair-collision contribution  to the pressure is thus:
\begin{equation}
\beta P^{\text{pair}}  = \frac{N}{V} \frac{N-1}{V} 2 \pi \sigma^2 g(2 \sigma), 
\end{equation}
an expression that is correct for finite $N$ and in periodic or \np 
boxes. 
The extrapolation of  $g(2 \sigma)$ from a histogram is detailed in
\app{app:Statistics}.
The range of distances to the wall that are eliminated is 
$[\sigma, \sigma (1 + \epsswell)]$, and the change in volume remains $2 
\epsswell V$. The contribution to the pressure of the wall at $x=0$ is then
\begin{equation}
\beta P^{\text{wall},-\ehatvec_x} = \frac{N \sigma \epsswell }{2 \epsswell V} 
\int \diff y \pi(\sigma, y) = \frac{N \sigma}{2 V L_x} \rhox(\sigma).
\end{equation}
where we used the rescaled line densities $\rhox$ and $\rhoy$ which remain 
\bigOb{1} for $V \to \infty$ (see \subsect{subsec:Chipping}). Summing 
over the four wall terms, one arrives at
\begin{equation}
\beta P^{\text{wall}} 
= \frac{N \sigma}{V } \glc \frac{\rhox(\sigma)}{L_x} + 
\frac{\rhoy(\sigma)}{L_y} \grc.
\end{equation}
The computation of the line densities $\rhox(\sigma)$ and  
$\rhoy(\sigma)$ from a histogram is detailed in \app{app:Statistics}.
The combined 
pair and wall contributions yield the homothetic pressure 
estimator for a \np $L_x \times L_y$ box:
\begin{subequations}
\begin{align}
&\beta P = N/V +  \nonumber \\
& \frac{N}{V}\glc2\pi\frac{(N-1)\sigma^2}{V} g(2\sigma) 
+ \sigma \frac{\rhox(\sigma)}{L_x} + 
\sigma \frac{\rhoy(\sigma)}  {L_y} \grc
 \label{equ:PressureSwellFiniteA}
\\
& \simasymp
 \frac{N}{V}\glc 1 + 2\eta g(2\sigma)\grc. 
 \label{equ:PressureSwellFiniteB}
\end{align}
\label{equ:PressureSwellFinite}%
\end{subequations}
\Eq{equ:PressureSwellFiniteA} is tested by histogram fits and extrapolations to 
contact of $\rho(\sigma)$ and $g(2\sigma)$  to $\fpn{4}{-3}$ in relative 
precision (see \tab{tab:UltraNaive}, line 6). \Eq{equ:PressureSwellFiniteB} 
has long been used for estimating pressures in MCMC~\cite{Metropolis1953}. 

EDMD and ECMC can estimate the pressure without the  extrapolations of the 
pair correlation functions and the wall densities by tracking the time
during which pairs of disks are close to contact, or a disk is close to the 
wall.
The explicit computation for EDMD simply reproduces 
\eqtwo{equ:FixLabelWK}{equ:MDswellCollision}, both 
at finite $N$ and in the thermodynamic limit. 
The corresponding homothetic pressure estimators for ECMC are readily derived, 
but they have diverging variances that require specific care. For all 
variants 
except straight ECMC, they correctly estimate wall contributions to the 
pressure and can be used for \np boxes. 
The velocities of straight ECMC are always parallel to some walls,  
precluding the estimation of all wall contribution to the pressure.

\subsubsection{Kinematic pressure estimators}
\label{subsec:KinematicEstimators}

Kinematic pressure estimators of EDMD determine the time-averaged exchange of 
momentum between disks, or between disks and a wall. Their use goes back to 
Daniel Bernoulli~\cite{Bernoulli1738}, who pointed out that under the scaling 
$\vvec_i \to \scalfac \vvec_i \forall i$ of \eq{equ:BernoulliScaling}, both the 
number of collisions per time interval and the momentum transmitted scaled as 
$\scalfac$, so that the pressure had to be proportional to the square of the 
(mean) velocity (in other words to the temperature). 
In the \np $L_x 
\times L_y$ box, the transmitted momentum with, say,  the vertical walls at 
$x=0$ and $x=L_x$ gives the kinematic EDMD estimator:
\begin{subequations}
 \begin{align}
 P_x &= \frac{1}{2L_y \tsim} \sum_{\wallcolls[,\pm \ehatvec_x]{i}} 2m 
 |\vperp_\text{wall}| 
 \label{equ:KinematicWallA} \\
     &= 2 m \mean{|\vperp_\text{wall}|} \nunitwallx 
\label{equ:KinematicWallB} \\
     &= \frac{m \RR \sqrt{\pi}}{N}\frac{\Gamma(N + \half)}{\Gamma(N)} 
     \nunitwallx, 
\label{equ:KinematicWallC}\\
	&\simasymp \sqrt{ 2 \pi \beta m} \nunitwallx \label{equ:KinematicWallD},
 \end{align}
\label{equ:KinematicWall}%
\end{subequations}
where in \eq{equ:KinematicWallC}, we used \eq{equ:MaxwellBoundaryB}. Already at 
finite $N$, the kinematic
EDMD estimator  of \eq{equ:KinematicWallC} is identical to the 
thermodynamic wall-rift pressure 
estimator of \eq{equ:PressureMDWallc}, as we may identify $ \RR^2 
= 2N / (m \beta)$. 

The EDMD kinetic pressure estimator can also be derived 
from the virial function
\begin{equation}
     G_x = m\sum_{i=1}^N  x_i  v_{i,x} 
\end{equation}
which is strictly bounded during molecular dynamics, so that  
its mean time derivative vanishes:
\begin{multline}
\mean{\frac{\diff}{\diff
 t} G_x}=  
\mean{\GDot[wall]{x}} +
\mean{\GDot[pair]{x}} +
m\mean{\sum_{i=1}^{N} v_{i,x}^2}\\
= m \mean{\sum_{i=1}^N   (  x_i \dot v_{i,x} + 
 v_{i,x}^2 )} = 0. 
\label{equ:deriv}
\end{multline}
The wall contribution to this expression stems from collisions with 
the vertical walls at $x = 0$ and $x = L_x$, which are given by
$2m\mean{|\vperp_{\text{wall}}|} \sigma$ 
and $-2m\mean{|\vperp_{\text{wall}}|}(L_x-\sigma)$, respectively. This results 
in 
\begin{subequations}
\begin{align}
\mean{ \GDot[wall]{x}} &= -2m\mean{|\vperpwall|}(L_x 
/ 2-\sigma) \nwallx \\
&= -V P_x  + 2 m \sigma \mean{ | \vperpwall | } 2 L_y \nhat^{\pm x 
}_{\text{wall}}, 
\end{align}
\end{subequations}
where we have used \eq{equ:KinematicWallB}.

For the pair-collision contribution to \eq{equ:deriv}, we use 
that at the collision of disks $i$ and $j$, the distance $\Delta \xvec_{ij} = 
\xvec_i - \xvec_j$ satisfies $| \Delta \xvec_{ij} | = 2 \sigma$. With the unit 
vector $\ehatvec_\perp = \Delta \xvec_{ij} / (2 \sigma)$ and the velocity 
difference $\Delta \vvec = \vvec_i - \vvec_j$ (before the collision), 
the change of the velocity  of disk $i$ is $- \ehatvec_\perp \scal{\Delta \vvec}
{\ehatvec_\perp} $, and 
the change of the velocity  of disk $j$ is $ \ehatvec_\perp \scal{\Delta 
\vvec}{ 
\ehatvec_\perp} $ (see~\cite[Sect. 2.1.1]{SMAC}).
An individual pair collision thus contributes
\begin{equation}
 -m \frac{(\delxminij)^2}{4\sigma^2}
\underbrace{\scal{ \Delta \vvec}{ \Delta \xvec_{ij}^\text{min}}}_{-2 \sigma 
|\vrelperppair|}, 
\end{equation} 
an expression where both terms can be averaged independently.
Finally, we may use $ \mean{v_{i,x}^2} =  \RR^2 /(2N)$
to rearrange \eq{equ:deriv}
into a kinematic EDMD pressure estimator:
\begin{multline}
\beta P_x = \frac{N}{V} +  
\frac{\beta}{V \tsim}
\sum_{\wallcolls[,\pm \ehatvec_x]{i}}\mean{2\sigma|\vperpwall| } \\
+\sum_{\paircolls} 
\frac{(\delxminij)^2}{4\sigma^2}
\mean{2 \sigma \vrelperppair}.
\label{equ:pressurefluxX}
\end{multline}
Using \eqtwo{equ:MaxwellBoundary}{equ:MaxwellPair}, 
this
kinematic estimator is seen to be equivalent to the thermodynamic 
rift-average estimator of \eq{equ:MDrift}.

\section{Equation-of-state computations}
\label{sec:EquationsOfState}

We now compare historic hard-disk pressure computations since 
1953~\cite{Metropolis1953,Zollweg1992,Jaster1999PRE,Jaster2004,Mak2006, 
Alder1962,Engel2013,QiGantaparaDijkstra_2014} with massive simulation results 
obtained in this work using the sampling algorithms and pressure estimators of 
\sect{sec:Algorithms}. Our re-evaluation will illustrate the three principal 
challenges that the hard-disk model shares, \emph{mutatis mutandis}, with other 
sampling problems. First, the estimate of the pressure  continues to depend on 
the initial configuration for very long run times, until equilibrium is 
reached. We will call this time the 
\quot{mixing time}~\cite{Levin2008} in a slight abuse of terminology, as we 
do not consider certain pathological initial configurations, which trap the 
Markov chain forever (see~\cite{Hoellmer2022}). 
The pioneering works, by Metropolis \etal\ and by 
Alder and Wainwright obtained crucial insights from very short 
computer experiments with run times much below the mixing time.
However, later works that attempted to interpret manifestly unequilibrated 
samples~\cite{WeberMarxBinder1995,MitusWeberMarx1997,Binder_2002}, or that 
failed to recognize the lack of convergence, arrived at qualitatively wrong 
conclusions. 
In the hard-disk model, 
mixing times can be bounded rigorously only at small 
densities~\cite{Kannanrapidmixing2003}.
At higher densities, 
heuristic criteria for 
the mixing time, which have not been fully 
presented before, appear crucial. 
In our case, they depend 
on time series of other observables than the pressure, or on multiple runs 
from qualitatively different initial configurations.
 
The second challenge for hard-disk computations consists in the  intricate
dependence of the pressure on the shape of the box (that is, the aspect 
ratio) and on the number $N$ of disks,
rendering extrapolations to the thermodynamic limit non-trivial. 
In small boxes, 
the  hexatic and solid phases are confounded, as they only differ at large 
distances, so that the behavior in the 
thermodynamic limit is not necessarily reflected in the equation of state at 
small $N$.

The third challenge concerns the very evaluation of the pressure.
Within Monte Carlo methods, the pressure was long
evaluated through extrapolation towards contact of the pair correlation
function in \eq{equ:PressureSwellFiniteB},
a procedure fraught with uncertainty.
The rift formulas of \subsect{subsec:PressureComputations} that originated with
ECMC, and that we even use in MPMC as short fictitious ECMC runs 
placed at regular time intervals, overcome the need
for extrapolations.

\subsection{Hard-disk equation of state for small $N$}
\label{subsec:EOS_Classics}

Since the early days of computer simulation, the pressure of the hard-disk 
model has been computed  with the aim of determining its phase behavior in the 
thermodynamic limit. While the identification of thermodynamic phases 
in finite systems can be subject to discussion, the 
pressure is unambiguously defined, and it can in principle be computed 
to arbitrary precision.

\begin{figure*}[htb]
\centering
\includegraphics[width=2.1\columnwidth]{
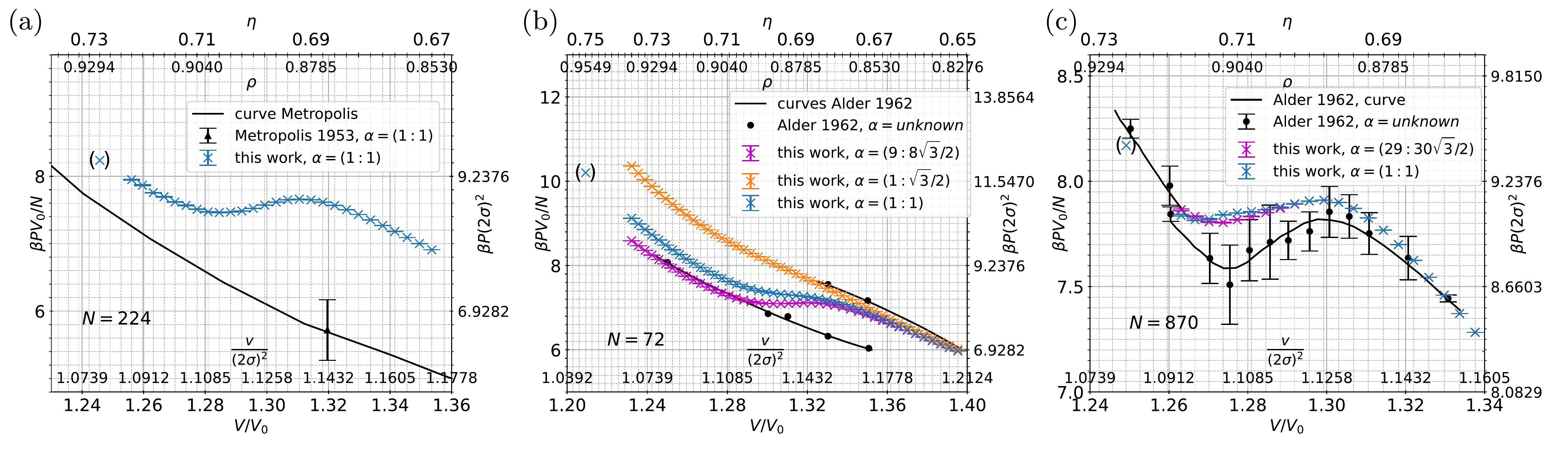}
\caption{
Equations of state $P(V)$ for the  hard-disk model at small $N$. 
\subcap{a} 
$P(V)$ for 
$N=244$ for $\aspect{1}{1}$
computed in 1953~\cite{Metropolis1953} compared with ECMC computations 
(this work).
\subcap{b} $P(V)$ for  $N=72$ computed in
1962~\cite{Alder1962} (for unspecified aspect ratio $\alpha$) and ECMC pressures
for $\aspect{1}{\sqrt{3}/2}$, $\aspectclean{1}{1}$, and $\aspectclean{9}{8
\sqrt{3}/2}$. \subcap{c} $P(V)$ for $N=870$. This work's square-box 
computations satisfy
$\mean{\Psi_6} \simeq 0$, except for data points in parentheses (see 
\fig{fig:Psi6Dependence}). 
}
\label{fig:SynopsisSmallSystems}
\end{figure*}

\subsubsection{Metropolis \etal\ (1953), rotation criterion}
\label{subsec:EOS_Classics2}
In $1953$, in the publication that first introduced MCMC, Metropolis 
\etal~\cite{Metropolis1953} estimated the pressure from the extrapolated 
pair-correlation function for $224$ disks in a periodic square box. This number 
 $224= 16 \times 14 $ disks can be perfectly packed 
at density $\eta =  \pi/ (2 \sqrt{3}) = 0.907$ in an almost square-shaped box of 
aspect ratio $\aspect{16 \sqrt{3}/2}{14} = \aspectclean{0.9897}{1}$ and almost 
at that density in a perfectly square box. Metropolis \etal\ concluded that 
\quot{(t)here is no 
indication of a phase transition}. The equation of state for $N=224, 
\aspect{1}{1}$, recomputed in this work using straight ECMC to a relative 
precision of $10^{-4}$, is somewhat higher than 
the historic pressures. It is also slightly non-monotonic
(see \subfig{fig:SynopsisSmallSystems}{a}). 

The $224$-disk square-box system of Metropolis \etal, from $1953$, carries 
lessons that are pertinent to the present day. Indeed, 
in a square box, 
any
hard-disk sample can be rotated by an angle $\pi/2 =
90$\textdegree\ into another valid sample. At high enough density, two 
such
samples are inequivalent because the
local hexagon which  describes the six disks that typically surround any
given disk has on average a $60$\textdegree\ symmetry but not a $90$\textdegree\
symmetry. For each local set of samples, there thus exists another inequivalent 
set of samples (generated through a $90$\textdegree\ rotation) of identical
statistical weight.
This rotation, and the corresponding transformation of samples  can be 
formalized through the global orientational order parameter
\begin{equation}
  \Psi_6 =  \frac{1}{N} \sum_l \frac{1}{\text{nbr}(l)}
  \sum_{j=1} ^{\text{nbr}(l)} e^{6 i \phi_{lj}},
\label{equ:OrientationalOrder}
\end{equation}
that changes from $\Psi_6$ to $-\Psi_6 $ (that is, $\arg( \Psi_6) \to
\arg( \Psi_6) + \pi$) under a rotation by $90$\textdegree. In
\eq{equ:OrientationalOrder}, $\phi_{lj}$ is the angle of the line connecting
disks $l$ and $j$ with respect to the $\ehatvec_x$-axis, and $\text{nbr}(l)$ is 
the
number of neighbors of $l$ resulting from a Voronoi construction.
In a square box, the ensemble
average of the orientational order parameter thus satisfies
$\mean{\Psi_6}=0$, and for an irreducible Markov chain, it agrees with its
time average, as expressed in the ergodic theorem~\cite{Levin2008}: 
\begin{equation}
\prob_{\pit{0}}
 \glc \lim_{t \to \infty}
 \frac{1}{t} \sum_{i=1}^t f_i
 = \mean{f} \grc = 1,
\label{equ:ErgodicTheorem}
\end{equation}
where $f$ is a function of the sample at time $i$ given the 
distribution $\pit{0}$ of initial configurations, and $\prob$ is the 
probability. 
As the mean value $\mean{\Psi_6}$ is known to 
vanish, we can employ \eq{equ:ErgodicTheorem} with $f = \Psi_6$ as a 
diagnostic tool and suppose that the hard-disk Markov chain in a square 
box reaches the mixing time (with errors decreasing as the square 
root of the run time)
only when 
the  orientational order parameter $\Psi_6$ has been rotated by more than 
$180$\textdegree~\footnote{In our simulations, we confirm that the 
orientational order parameter $\Psi_6$ has been rotated more than 
$90$\textdegree~and has visited at least one of the two points on the real axis, 
$\Psi_6 \simeq \pm |\Psi_6|$, and one on the imaginary axis, $\Psi_6 \simeq \pm 
i|\Psi_6|$.}:
\begin{equation}
 |\text{supp}(\arg(\Psi_6) | >  \pi.
 \label{equ:RotationCriterion}
\end{equation}
Here, \quot{$\text{supp}$} stands for the support of the empirical 
distribution, 
in our case for the range of angles of $\Psi_6$  that are visited during a 
simulation.
This heuristic rotation criterion supposes that the orientational order 
parameter $\Psi_6$ is the slowest-decaying variable in the hard-disk system. 
Our time series of $\Psi_6$ in a square box, with known mean value 
$\mean{\Psi_6} =0$, 
pinpoints problems with a hard-disk pressure estimation that might not 
be signaled by the time series of the pressure itself. We use the 
rotation criterion 
in two different settings. In small systems (as 
the $224$-disk case of Metropolis \etal), the entire range of $\arg(\Psi_6)$ 
values is swept through many times, leading to high-precision estimates for the 
pressure, even though it strongly depends on the angle. In large 
systems, as the 
hard-disk 
model with $N=128^2$ at $\eta=0.716$, that we will discuss in  
\fig{fig:PostAnalysis128}, we barely satisfy the criterion, but it still 
assures 
us that up to a symmetry $\Psi_6 \to -\Psi_6$
all relevant regions of sample space were visited. High-precision 
estimates for the pressure now result from the fact that the pressure 
depends weakly on $\arg(\Psi_6)$.

Our ECMC simulations satisfy the rotation criterion for $N=224$ in a periodic 
square box up to a density $\eta =0.72$. At large enough densities, the ECMC 
simulation may remain for long times in a set of samples with essentially the 
same value of $\arg(\Psi_6)$ before flipping to another set of samples with 
$\arg(\Psi_6) + \pi$. This very slow rotation of $\Psi_6$ is a harbinger of the 
serious convergence problems of the hard-disk model for larger $N$ at densities 
of physical interest.  As we will show, the pressure is strongly correlated with 
$\Psi_6$ up to moderate values of $N$. 

\subsubsection{Revisiting Alder and Wainwright (1962)}
\label{subsec:EOS_AlderWainwright}
Alder and Wainwright, in $1962$, used EDMD to estimate the pressure  for $72$ 
and  $870$ disks in rectangular periodic boxes for which they did not specify 
the aspect ratios. As already discussed in \fig{fig:AlderWainwrightRevisited}, 
their non-monotonic equation of state led to the prediction of a phase 
transition. The computed pressure is independent 
of the sampling method (molecular dynamics, local Metropolis algorithm, ECMC), 
but it depends on the  aspect ratio of the box. For $72=9\times8$ disks and 
aspect ratio $\aspect{9}{8\sqrt{3}/2}  = \aspectclean{1}{0.7698} $ where they 
can be perfectly packed, the equation of state obtained by ECMC in this work 
agrees remarkably well 
with the historic data (see 
\subfig{fig:SynopsisSmallSystems}{b}). In contrast, 
for a square box (aspect ratio $\aspect{1}{1}$), the equation of state follows 
a slight \quot{S} shape, but it remains monotonous for all densities. For the 
aspect ratio $\aspect{1}{\sqrt{3}/2}$, the pressure is barely \quot{S} shaped. 
For the aspect ratio $\aspect{1}{1}$, our ECMC computations satisfy the 
rotation criterion up to densities $\eta \lesssim 0.74$, and pressure estimates 
achieve $10^{-4}$ relative precision.

\begin{figure}[htb]
\centering
\includegraphics[width=\columnwidth]{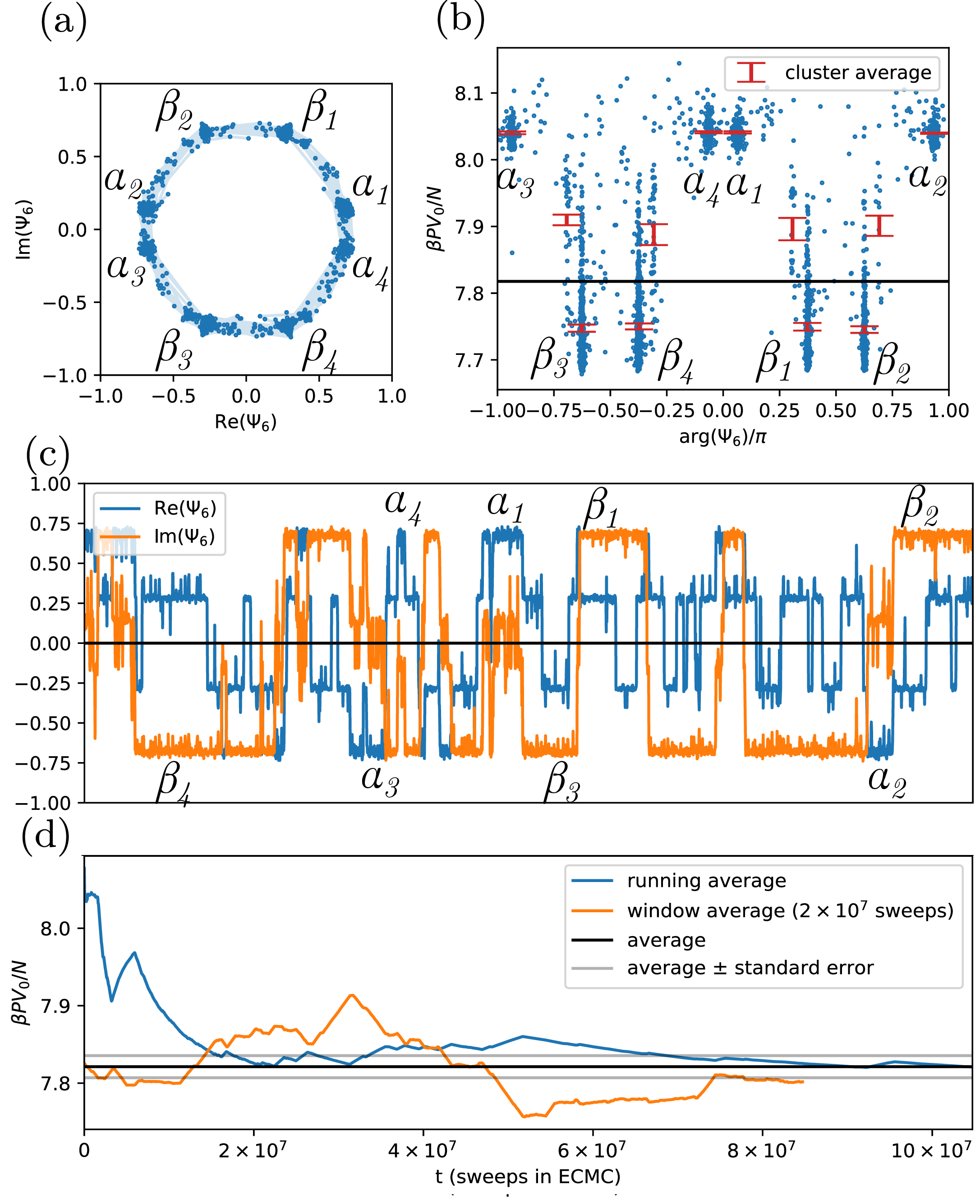}
\caption{Pressure $P$ and global orientational order $\Psi_6$ for a three-hour
ECMC run ($N=870$, $\eta = 0.716$, square box $\aspect{1}{1}$). \subcap{a}
Values of $\Psi_6$
in the complex plane. Highlighted clusters with
inverted
$\Psi_6$ (such as $\alpha_1$ and $\alpha_3$) have the same statistical weight.
\subcap{b} Cluster averages for $P$ \vs\ $\arg(\Psi_6)$.
\subcap{c} Trajectories of $\text{Im}(\Psi_6)$ and $\text{Re}
(\Psi_6)$ with indicated clusters. \subcap{d} Running
average and window average for $P$.}
\label{fig:Psi6Dependence}
\end{figure}

For $870$ disks, the dependence of the  pressure 
on the aspect ratio is less pronounced than for $N=72$
(see \fig{fig:SynopsisSmallSystems}{c}).
Since $870 = 30\times 29$, this
number of disks can be close-packed for the aspect ratio $\aspect{29}{30
\sqrt{3}/2} = \aspectclean{1}{0.896}$. For the aspect ratio $\aspect{1}{1}$,
the orientation criterion of \eq{equ:RotationCriterion} is again satisfied up
to high densities (see \subfig{fig:Psi6Dependence}{a}). However, even at 
moderate densities, an ECMC run can take 
several CPU hours before visiting all possible orientations, and the 
pressure clearly
correlates with the orientation (see \subfig{fig:Psi6Dependence}{b}). On 
smaller time scales, the time series remains blocked in samples that all 
roughly have the same orientational order parameter
$\Psi_6$ (see \subfig{fig:Psi6Dependence}{c}). Analyzing such shorter time 
series gives incorrect estimates of the pressure ($P_\alpha $ or $P_\beta$, 
\etc, rather than $P$).
Accordingly, 
the window-averaged pressure features  long-time correlations, with an estimated
autocorrelation time of  $\sim \fpn{2}{10}$ events (corresponding to roughly 
two CPU hours for ECMC). Nevertheless, on a long enough time scale estimated by 
the rotation criterion, all these systematic errors disappear, and the 
error of the pressure estimator starts to decrease as the square root of the 
run time (see \subfig{fig:Psi6Dependence}{d}). The achieved $10^{-4}$ relative
error on the pressure estimates in 
\subfig{fig:SynopsisSmallSystems}{c}, from longer simulations than those 
illustrated in \fig{fig:Psi6Dependence}, is much smaller than the
systematic error $|\mean{P_{\alpha}} - \mean{P_{\gamma}}| / P \sim
10^{-2}$ of a calculation that is too short to rotate $\Psi_6$.  

\begin{figure*}[htb]
\centering
\includegraphics[width=2.1\columnwidth]{
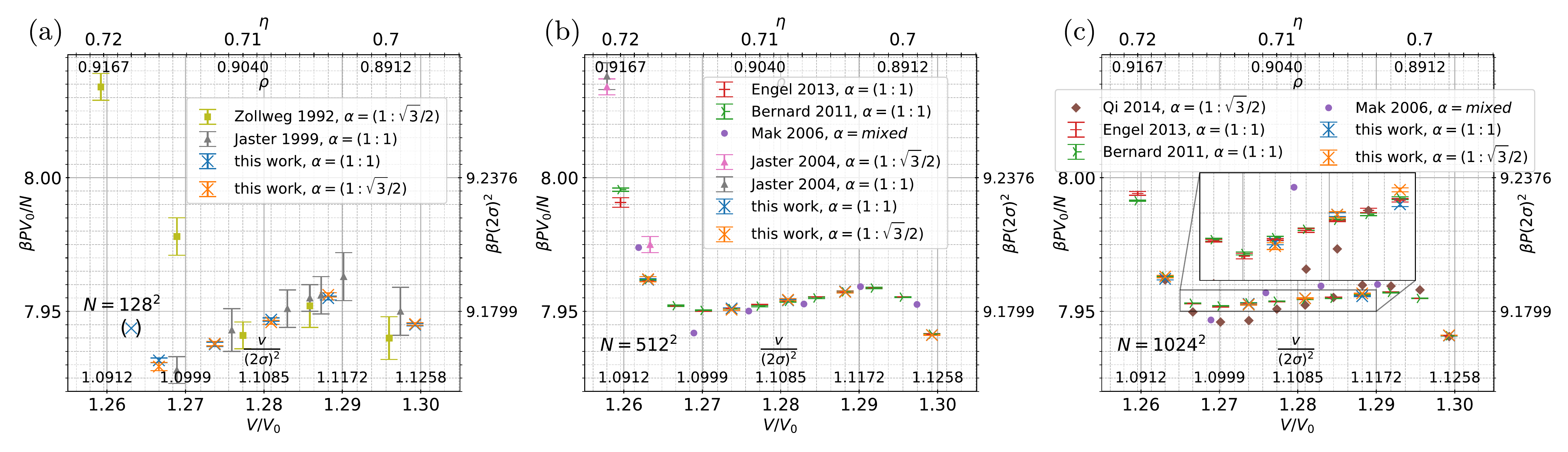}
\caption{ Equations of state $P(V)$ for the hard-disk model at large $N$. 
\subcap{a} $P(V)$ 
for $N=128^2$ from Refs~\cite{Zollweg1992,Jaster1999} and MPMC pressures (this
work)  for $\aspect{1}{\sqrt{3}/2}$ and
$\aspect{1}{1}$
where
all but the data point in parentheses satisfy the rotation 
criterion of
\eq{equ:RotationCriterion} 
(see \fig{fig:PostAnalysis128} for $\Psi_6$-resolved pressures).
\subcap{b}  $P(V)$  for $N=512^2$ from
Refs~\cite{Jaster2004,Mak2006,Mak2006,Engel2013} and MPMC pressures (this work)
for aspect ratios $\aspect{1}{\sqrt{3}/2}$ and $\aspect{1}{1}$, where 
runs
with $\eta < 0.712$  satisfy the rotation criterion.
\subcap{c} $P(V)$ for $1024^2$ from
Refs~\cite{Mak2006,Bernard2011,Engel2013,QiGantaparaDijkstra_2014}, 
compared to MPMC (this work)
for aspect ratios $\aspect{1}{\sqrt{3}/2}$ and $\aspectclean{1}{1}$, where at
density $\eta > 0.708 $ the rotation criterion is violated, but the systematic
error thus committed is negligible (see \subfig{fig:XXLAnalysis}{a}).}
\label{fig:SynopsisLargeSystems}
\end{figure*}

\subsection{Equations of state for large $N$}

The equations of state for larger $N$ than those considered by Metropolis \etal\ 
and by Alder and Wainwright came into focus in the decades following 1962. At 
sufficiently large $N$, as we know today, fluid, hexatic, and solid phases can 
be distinguished, and the latter clearly differs from the crystal. In the 
simulations, three effects stand out. First, mixing and 
autocorrelation times 
become truly gigantic already for reasonable densities, even for the best 
currently known algorithms. Nevertheless, the pressure (as other physical 
quantities that we do not consider in this work) can be computed to a precision 
that, from a given time on, increases as the square root of the computer time. 
From $N=128^2$ to $N= 1024^2$, this program can be put into practice, but it 
requires considerable computer resources. The failure to converge is signaled 
through a number of criteria, but not necessarily by the time 
series itself. Second, in the coexistence phase of the fluid and the hexatic in 
the $NVT$ ensemble, the initial dynamics towards equilibrium is dominated by 
coarsening. In this process, for example, small hexatic islands nucleate in the 
fluid, then coalesce and slowly grow until, in the stationary 
state of the time evolution, the sample presents itself as two domains, one for 
each coexisting phase. Precise knowledge of the pressure allows one to draw the 
boundaries of the phase coexistence. Third, as realized ten years ago, the 
high-density coexisting phase through the first-order phase transition is a 
hexatic, and thus distinct from the crystal that can serve as an initial 
configuration of MCMC configurations. Mixing times in the  hexatic phase are 
very long, and are likely to scale with a larger exponent with $N$ than in the 
fluid~\cite{Helmuth2022}.

\subsubsection{Hard-disk model with $N=128^2 $ to $N=512^2$}

\begin{figure}[htb]
\centering
\includegraphics[width=\columnwidth]{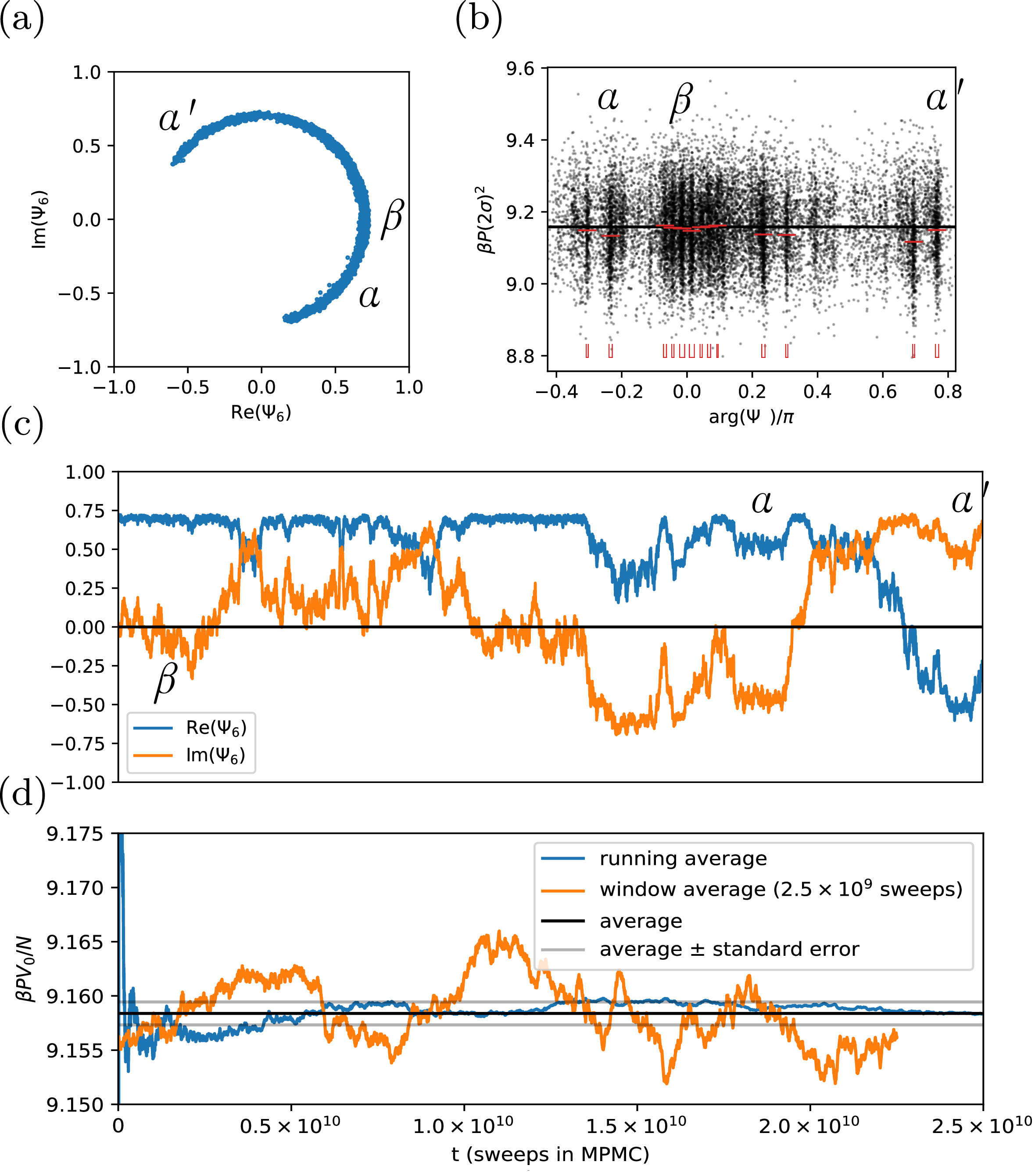}
\caption{Analysis of the rotation criterion of \eq{equ:RotationCriterion}
for a single MPMC run ($N= 128^2$, $\eta=0.716$, $\aspect{1}{1}$).
\subcap{a} Histogram of $\Psi_6$ in the complex plane.
Clusters $\alpha$, $\alpha'$ satisfies $\Psi_6^{\alpha} =
-\Psi_6^{\alpha'}$ and have equal weight. \subcap{b} Trajectory of
$\text{Re}(\Psi_6)$ and $\text{Im}({\Psi_6})$
with first visits to clusters indicated (\cf\ \fig{fig:Psi6Dependence}).
\subcap{c} Running average and window average of the pressure, showcasing
slow convergence.}
\label{fig:PostAnalysis128}
\end{figure}

For the hard-disk model with $N=128^2$, the relative precision levels of 
sequential ECMC, parallel ECMC, and of MPMC reach $\sim 10^{-5}$ for the 
pressure, for example, at $\eta = 0.698$, much more precise than previous 
studies in the literature~\cite{Zollweg1992,Jaster1999} (see 
\subfig{fig:SynopsisLargeSystems}{a}). For $\aspect{1}{1}$, our calculations 
satisfy the rotation criterion of \eq{equ:RotationCriterion} up to density $\eta 
= 0.716$, albeit for high densities on an impressive time scale, even for the 
MPMC algorithm (see \fig{fig:PostAnalysis128}). This is where
earlier studies failed to equilibrate, and produced erroneous 
pressure estimates.

At density $\eta=0.716$, samples of the MPMC computation may remain in one 
cluster indexed by a given value of $\arg(\Psi_6)$ for $\sim 10^{9}$ sweeps, and 
then produce a cluster average for the pressure $\beta P V_0 /N$ that differs 
relatively by about $10^{-3}$ from the equilibrium average (see 
\subfig{fig:PostAnalysis128}{c}). On such time scales, the outcome of the 
simulation is thus unpredictable, and the observed convergence of the pressure 
is not towards its ensemble average but towards some metastable cluster value. 
This behavior is readily detected from within the simulation data through the 
rotation criterion of \eq{equ:RotationCriterion} and through the dependence of 
obtained pressure values on initial conditions (such as different orientations 
or fluid and crystalline initial configurations).

For even larger systems, such as $N=512^2$, the computations in the literature 
dramatically suffer from the failure to equilibrate, with incorrect pressure 
estimates especially at high densities. For $\aspect{1}{1}$, our MPMC 
implementation satisfies the rotation criterion at 
$\eta \lesssim 0.712$ within a few weeks of computer time (which would 
correspond to centuries of run time of the local Metropolis algorithm on a 
single CPU). For even higher 
densities, all currently known sampling algorithm fail to equilibrate in the 
strict sense of that criterion.
Fortunately, at larger $N$, the influence of the boundary conditions is much 
smaller than for small $N$ (see \subfig{fig:XXLAnalysis}{a}). We estimate the 
systematic error stemming from the failure to rotate $\Psi_6$ by starting 
independent simulations for $N=512^2$ from a number of initial configurations 
with different 
global orientational order parameters $\Psi_6$ (see 
\subfig{fig:XXLAnalysis}{a}). The resulting systematic errors are found to be 
at most as large as the 
statistical errors. Our pressure values are consistent with 
previous ECMC and MPMC calculations~\cite{Bernard2011,Engel2013} up to $\eta = 
0.718$, cross-validating the correctness of the conclusion in 
Ref.~\cite{Engel2013} (see \subfig{fig:SynopsisLargeSystems}{b}). In a 
non-square box, the components $P_x$ and $P_y$  of the pressure generally  
differ. For $N = 128^2$ and $512^2$ at aspect ratio $\aspect{1}{\sqrt{3}/2}$, 
our estimates for $P_x$ and $P_y$ agree within error bars even in the hexatic 
phase, as the system dimensions are larger than the positional correlation 
length. 

\subsubsection{Hard-disk model at $N=1024^2$}
\label{subsec:LargeSystems}

For the hard-disk model at $N=1024^2$, single-core
implementations of the reversible Metropolis algorithm and of EDMD
fail to equilibrate for densities  $\eta\gtrsim 0.700$ on accessible time 
scales 
even on a modern CPU. 
Only straight ECMC (whose week-long mixing time of the serial
version~\cite{Bernard2011} reduces in the parallel implementation) and MPMC 
(run in parallel on thousands of cores on a GPU) are currently able to partially 
achieve convergence. It is for this reason that in the past, unconverged 
calculations~\cite{Jaster2004,Mak2006,QiGantaparaDijkstra_2014} 
resulted in erroneous pressure estimates and,
in consequence, qualitatively wrong predictions for the hard-disk phases and 
the phase transitions.

\begin{figure*}[htb]
\centering
\includegraphics[width=2\columnwidth]{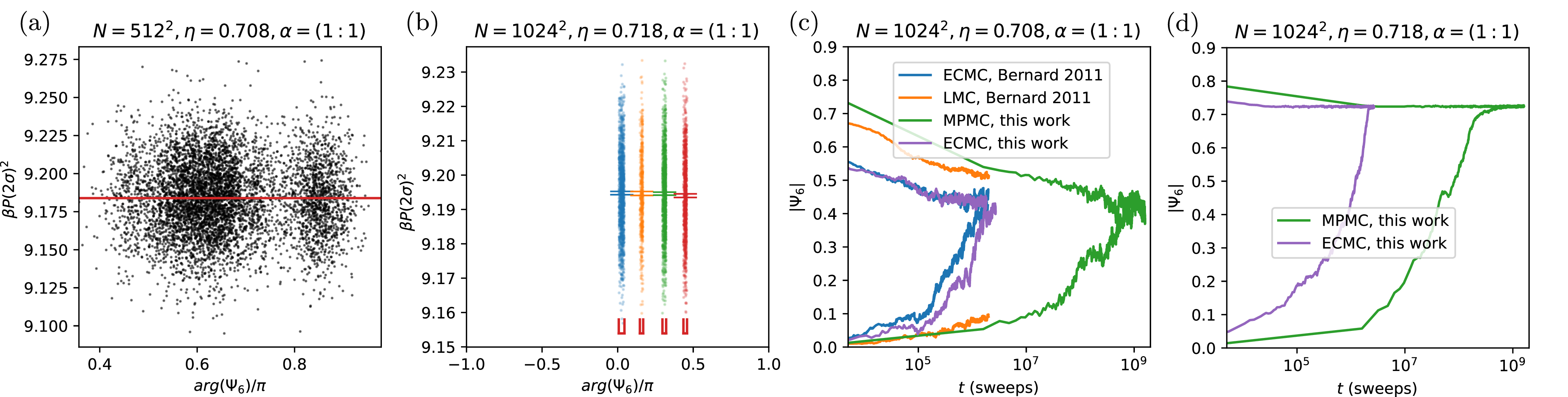}
\caption{Convergence analysis for the hard-disk model at 
$\eta=0.708$ and $\eta= 0.718$ (square box $\aspect{1}{1}$).
\subcap{a} Scatter plot of the pressure as a function of the orientational 
order parameter ($N=512^2$).
\subcap{b-d} $N=1024^2$ for EDMD and ECMC.
\subcap{b} Cluster averages obtained from independent runs
from initial
configurations at specific values of $\Psi_6$. The difference, smaller than
$10^{-3}$, estimates the systematic error.
\subcaptwo{c}{d} Time evolution of the absolute orientational order
$|\Psi_6|$, starting from either a
disordered initial configuration (with $|\Psi_6| \gtrsim 0$)
or from a crystal (with $|\Psi_6| \lesssim 1$)
(LMC refers to the local Metropolis algorithm).
}
\label{fig:XXLAnalysis}
\end{figure*}

The slow mixing manifests itself in pairs of runs that start on the one hand 
from a 
fluid-like initial configuration with only short-range correlations and a global
orientational order parameter $|\Psi_6| \gtrsim 0$ (obtained by the
Lubachevsky--Stillinger algorithm~\cite{Lubachevsky}) and on the other hand
from a crystalline initial configuration with $|\Psi_6| \lesssim 1$. In the 
fluid--hexatic coexistence region ($\eta = 0.708$), as well as in the 
hexatic phase ($\eta = 0.718$), ECMC takes about 
$10^6 $ sweeps to coalesce the two 
values of $|\Psi_6|$ (see \subfig{fig:XXLAnalysis}{c} and 
\subfig{fig:XXLAnalysis}{d}). For ECMC, at $\sim 10^{10}$ events/hour, this 
corresponds to about a week of single-core CPU time. In 
contrast, 
MPMC requires roughly $10^9$ sweeps to coalesce. On a GPU with $\sim 10^4$
individual cores, this is achieved in less than two days, but on a single-core
CPU, the local Metropolis algorithm (which has roughly the same efficiency per
move as MPMC) would require $10^{9+6}$ moves which correspond to $ \sim 10^5$
hours or $\sim 10$ years, at a typical $10^{10}$ moves per hour. 
Both branches of these calculations have similar times for arriving
at equilibrium, illustrating that the fluid--hexatic coexistence phase is as
difficult to reach from the fluid as it is from the crystal. While the mixing is 
very slow, the pairs of curves reaching 
the same value of $|\Psi_6|$ give a lower bound for the required run times of  
our ECMC and MPMC algorithms, although these times are still much below the 
mixing time in this system, if one were to include the rotation in 
$\arg(\Psi_6)$ in its definition. For the density $\eta = 0.718$, 
at $N=1024^2$, our total MPMC run times amount to $\fpn{6.4}{9}$ sweeps, 
roughly $6$ times longer than what it shown in \subfig{fig:XXLAnalysis}{d}.

Although MPMC and ECMC are today's fastest algorithms for the 
hard-disk model, they fail to satisfy 
the rotation criterion of 
\eq{equ:RotationCriterion} on human timescales for 
$N=1024^2$ at densities $\eta \gtrsim 0.708$. Fortunately, the influence 
of 
$\arg(\Psi_6)$ on the pressure is 
quite small. To test this, we started very long MPMC calculations from a number 
of finely spaced crystalline initial configurations with different values of
$\text{arg} (\Psi_6)$. 
At the very high density of $\eta=0.718$ for $N=1024^2$, 
the relative 
statistical errors for the pressure is $\fpn{5}{-4}$
for each run, while the maximum distance between the mean values, that could 
possibly correspond to a systematic error, is also found to be $\fpn{5}{-4}$.
We estimate the pressure uncertainty as the maximum of the individual 
statistical and the difference in mean values
(see \subfig{fig:XXLAnalysis}{b}). The 
estimated pressures are also almost independent of the aspect ratio 
$\aspect{1}{1}$ and the $\aspect{1}{\sqrt{3}/2}$ 
(\subfig{fig:SynopsisLargeSystems}{c}). Finally, in non-square boxes, the 
estimates for $P_x$ and $P_y$ agree to very high precision for large $N$, while 
they differ markedly in smaller systems (see \tab{tab:StateOfArt}).
The disagreement of previous 
calculations appears thus
rooted in the very long times to reach the correct values of $|\Psi_6|$.

\begin{table}[ht]
\centering
\begin{tabular}{llll}
\hline
$N$ & $\alpha$ & $P$ / $P_x,P_y$ &Method\\ 
\hline
$64$&$\aspectclean{1}{\sqrt{3}/2}$ & $8.065(3), 8.137(4)$ & Naive\\
& & $8.0671(9), 8.1402(9)$ & ECMC\\
$72$&$\aspectclean{9}{4\sqrt{3}}$ & $8.1614(4), 8.2382(5)$ & Naive \\
& & $8.1617(7), 8.2386(8)$ & ECMC\\
$256^2$&$\aspectclean{1}{1}$  & $9.172(5)$ & ECMC ($g(2\sigma)$) \\
& & $9.1707(2)$ & ECMC\\
& & $9.1708(1)$ & MPMC \\
$256^2$&$\aspectclean{1}{\sqrt{3}/2}$ & $9.176(6)$ & ECMC ($g(2\sigma)$)\\
& & $9.1703(2), 9.1704(3)$ &ECMC\\
& & $9.1704(1), 9.1705(1)$ &MPMC \\
$512^2$&$\aspectclean{1}{1}$ & $9.170(2)$ & ECMC ($g(2\sigma)$) \\
& & $9.1699(2)$ & ECMC\\
& & $9.1696(1)$ & MPMC\\
$512^2$&$\aspectclean{1}{\sqrt{3}/2}$ & $9.167(3)$ & ECMC ($g(2\sigma)$) \\
& & $9.1694(3), 9.1694(3)$ & ECMC \\
& & $9.1695(2), 9.1697(2)$ & MPMC \\
\hline
\end{tabular}
\caption{Cross validation of pressure estimates between straight ECMC (naive 
and state-of-the-art) and MPMC in periodic boxes of given aspect ratio 
$\alpha$, 
all at density $\eta=0.698$. MPMC integrates short fictitious runs of 
straight ECMC, in order to estimate pressures through \eq{equ:RiftPressure}. 
ECMC uses the rift-average estimator of \eq{equ:RiftPressure}, except where 
indicated to test
agreement of  the pair-correlation formula of \eq{equ:PressureSwellFiniteB}.
}
\label{tab:StateOfArt}
\end{table}

\section{Conclusion}
\label{sec:Conclusion}

In this work, we have discussed the  hard-disk pressure,
which was estimated in the the very first MCMC
computation in 1953, and in one of the earliest
molecular-dynamics computations, in 1962. We have 
argued that the difficulty of the pressure estimation had  not been fully 
realized in the decades-long controversy over the phase-transition scenario of 
this simple model. Our first aim was to provide the context for this 
computation through a discussion of the physics of the hard-disk model, of the 
sampling algorithms and pressure estimators and, crucially, of the criteria for 
bounding mixing times. Our second aim was to finally provide definite 
high-precision estimates 
of the pressure through massive computations and to compare them to the values 
from the literature, thereby ending a long period of uncertainty and doubt. In 
doing so, we hope to provide benchmarks for the next generation of sampling 
algorithms, estimators, and physical theories.

The history of the hard-disk model epitomizes a number of prime computational 
issues. One of them is the role of so-called \quot{computer experiments}, that 
is, of heuristic simulations which run for much less than the mixing time. 
The pioneering work of Alder and Wainwright was clearly of that type, as 
their published pressures explicitly depend on the initial configurations.

Computer experiments below the mixing-time scale are akin to 
perturbation expansions in the theory of liquids or in quantum 
mechanics, as the sampling below the mixing-time scale merely \quot{perturbs} 
around the crystalline or fluid initial configurations. Just like perturbation 
theory, 
such computer experiments can provide important insights, yet 
they have 
limited predictive power, as was evidenced by the 
decades-long 
controversy about the hard-disk phase transitions. 
Beyond the mixing time, the 
influence  of the initial configuration fades away exponentially, and 
exponential
convergence towards the equilibrium distribution sets in. Only the 
statistical errors remain. In this regime, MCMC and molecular-dynamics sampling 
unfolds
all its power. Although
the mixing and correlation time scales can be gigantic, as discussed, the goal 
of sampling beyond the mixing time scale must not be lost sight of. 

A crucial computational issue for MCMC and molecular-dynamics algorithms 
consists in estimating the mixing times. We have insisted 
that simply analyzing a time series (in our case, that of the 
pressure) is usually not sufficient. Furthermore, we have discussed 
two strategies to estimate these times reliably for the hard-disk model. First, 
we designed an observable---the orientational order parameter for hard disks in 
a square box---with known mean value. We then argued that as long as the 
run-time average of this observable differed considerably from its known mean, 
the mixing-time scale has  not yet been reached. Used for more than a 
decade~\cite{Bernard2009,Bernard2011}, this rotation criterion  supposes that 
the orientational order is the slowest-moving observable in the hard-disk 
system.

Our second strategy to
lend credibility to our MCMC calculations consists in 
starting from widely different initial configurations, following in the 
footsteps 
of Alder and Wainwright, yet accepting the result of the calculation only if 
the influence of the initial configuration has faded away. This approach
is related to the coupling approach for Markov 
chains~\cite{ProppWilson1996}. The fluid 
and crystalline initial configurations that we used to initialize Markov chains 
for 
$10^6 $ disks in the hexatic phase stand in for 
the worst-case initializations, 
as they are called for 
in the definition of the mixing time~\cite{Levin2008}. 
The mixing time 
provides the relevant time scale for analyzing MCMC calculations, and certainly 
the one where run-time averages become independent of how the Markov chain is 
initialized.

Finally, we emphasize the role of algorithm development, and of hardware 
implementations, even in the simple model of hard disks. 
In this work, we relied heavily on
ECMC, which, as evidenced in \subfig{fig:XXLAnalysis}{c and d} and d, speeds up 
MCMC simulations by several 
orders of magnitude. ECMC is a family of non-reversible Markov chains, rather 
than a specific algorithm, and variants of the original straight and 
reflective ECMC continue to be developed.
The opportunities granted by non-reversible Markov chains (and by 
MCMC approaches in general), are certainly very far from having all been 
explored. The recent extension of ECMC to 
arbitrary interaction potentials~\cite{Michel2014JCP} and in particular to the 
field of molecular simulation~\cite{Faulkner2018,Hoellmer2020}, carries 
considerable promise. The spectacular development of GPU hardware 
over the last fifteen years has greatly democratized parallel computations 
with, again, one of the cleanest applications being the hard-disk model. 
Decidedly, this simple model is a \quot{Drosophila} of statistical physics.

\begin{acknowledgments}
{P.H.~acknowledges
support from the Studienstiftung des deutschen Volkes and from Institut
Philippe Meyer. W.K.~acknowledges support from the Alexander von Humboldt
Foundation. We thank R.E. Kohler for helpful correspondence.}
\end{acknowledgments}

\appendix

\section{Extrapolation and statistics}
\label{app:Statistics}
Sampling algorithms output time series of configurations and of pressures 
(for example one value of the estimated $P_x $ for each event chain in $\pm 
\ehatvec_x$). Further analysis transforms this raw output into the pressure 
estimates and confidence intervals  provided with this work. The 
pressure estimators that rely on the extrapolation of pair-correlation 
functions and wall densities have been superseded in recent years by  the 
rift-average estimators. We nevertheless describe them here in order
to illustrate that the new estimators are perfectly sound. We also sketch 
the stationary-bootstrap method which estimates
the confidence intervals of the pressure time series.

\subsection{Extrapolation of pair correlations and wall densities}

The pressure estimator of \eq{equ:PressureSwellFiniteA} extrapolates the 
rescaled line densities 
$\rho_{x}(x)$ and $\rho_{y}(y)$ to $x= \sigma, L_x - \sigma$ and $y= \sigma, 
L_y -\sigma$, 
respectively, and the     pair-correlation function  $g(r)$ to contact at 
$r=2\sigma$. We use the fourth-order polynomial histogram fitting procedure of 
Ref~\cite{Bernard2011} contained in the \HD software package (see 
\app{app:DataCodes}). Within our MPMC production runs, however, we use 
the parameter-free rift-average estimators from 
fictitious straight-ECMC runs to estimate $P_x$ and $P_y$, rather than   
the extrapolation method.

To determine the rescaled line density $\rho_x(\sigma)$ (and similarly 
$\rho_y$), the 
$x$-coordinate 
of a disk at position $\xvec_i = (x_i, y_i)$ 
is retained in a histogram of bin size $10^{-3}\sigma$
if $x_i < 1.1\sigma$ or $x_i  > L_{x} -1.1 \sigma$. 
The histogram is normalized by dividing the number of elements in each bin by 
$2 \times 10^{-3}\sigma n N$, where $n$ is the total number of sampled 
configurations 
(not only those contributing to the histogram) and $N$ 
is the number of disks. The histogram is further multiplied by $L_x$ for 
$\rho_x$ (and likewise by $L_y$ for $\rho_y$) in order to satisfy the 
normalization $\piOne(\xvec_i) = \rho(\xvec_i)/V$.
It is the line density $\rhox(x)$, which is 
fitted  and then extrapolated to $x=\sigma$.

The extrapolation of $g(2\sigma)$ proceeds analogously to that of
$\rho_{x}(x)$. The 
pair distances 
in the range  $ 2 \sigma < r < 2.1\sigma$ 
are retained in a histogram, then 
normalized by dividing the number of elements in each bin by 
the bin size $10^{-3}\sigma$ and the total number of sampled distances $n N(N - 
1)/2$. The normalized histogram approximates  $2\pi r \ghat(r)$. The 
histogram is further multiplied by $V/2\pi$ and divided by the distance 
$r$ corresponding to the center of each bin, yielding the empirical $g(r)$, 
that is then extrapolated to $r=2\sigma$. 

\subsection{Statistics}
The standard errors  in this work were computed with the 
stationary bootstrap method~\cite{Politis1994, nishikawa2021stationary}, and 
double-checked with the blocking method~\cite{Flyvbjerg1989}. In 
stationary bootstrap, the standard error is estimated by creating a large 
number of simulated time series (typically $1000$). 
Each of the time series has the same length as 
the original series, and is created by piecing together randomly chosen 
sub-series of geometrically distributed length. The only parameter controlling 
the
geometrical distribution is chosen so that it minimizes 
the mean squared error of the estimated standard error for an infinite 
sub-series length and for an infinite number of 
sub-series~\cite{Politis2004,Patton2009}. 
The compatibility of the stationary-bootstrap error estimate 
with that of the blocking 
method was carefully checked for the entire data presented in the figures 
as well as in \tab{tab:StateOfArt}.

\section{Historic data, codes, and validation}
\label{app:DataCodes}

The present work is accompanied by the \HD data and software package, which is 
published as an open-source project under the GNU GPLv3 license. \HD is 
available on GitHub as part of the \texttt{JeLLyFysh} organization~\footnote{The 
url of the repository is \url{https://github.com/jellyfysh/HistoricDisks}.}. The 
package provides the pressure data extracted from the literature since $1953$, 
and also the set of  high-precision pressures  of the present work (see 
\subsect{app:DigitizedPressures}). Furthermore, the package contains naive MCMC 
and MD implementations and pressure estimators 
(used for validation purposes in \tab{tab:UltraNaive}) as well as 
state-of-the-art implementations used in \sect{sec:EquationsOfState}.

\subsection{Pressure data, equations of states}
\label{app:DigitizedPressures}

The pressure data in the \HD package are from 
Refs~\cite{Metropolis1953, Alder1962, Zollweg1992, Jaster1999, 
Jaster2004, Mak2006, 
Bernard2011,QiGantaparaDijkstra_2014}, or else correspond to results
obtained in this work. Pressure data for a given reference, a 
given system size and aspect ratio are stored in a separate file in the 
\texttt{.csv} format (see the \texttt{README} file for details). Pressures and 
error
bars were digitized using the WebPlotDigitizer software~\cite{Rohatgi2020} 
where applicable, or else extracted from published tables.
The \HD package furthermore provides Python programs that visualize
equations of state.
All pressure data are for the $NVT$ ensemble, and the control
variable (volume or density, plotted on the $x$-axis)
follows all four conventions of
\eq{equ:DensityVolume}.
The dependent variable (the pressure, plotted on the $y$-axis)
follows two conventions, namely $\beta P V_0 / N$ and
$\beta P (2\sigma)^2$.
In order to
facilitate
the direct comparison  across different conventions, the produced figures have
four $x$-axes and two $y$-axes. The pressure data base in the \HD package may 
evolve in the future.

\subsection{Computer programs}
\label{app:AccessComputer}

In addition to pressure data, the \HD{} package provides access to sampling 
algorithms (local Metropolis algorithm, EDMD, and several variants of ECMC). 
Each algorithm is implemented in two versions. A \ultranaive version for four 
disks in a \np rectangular box is patterned after Ref.~\cite{SMAC}. A naive 
version for $N$ disks in a periodic rectangular box is useful for validation 
of more advanced methods. Both versions are implemented in Python3 (compatible 
with PyPy3). In addition, the package provides  a state-of-the-art ECMC program 
for hard disks. 

\subsubsection{Four-disk \np-box programs}
\label{subsec:FourDisksInSquare}

Our naive programs consider four disks of radius $\sigma=0.15$
in a \np square box of sides $1$.
We implement the
Metropolis algorithm,  EDMD, and the straight,
reflective~\cite{Bernard2009}, forward~\cite{Michel2020} and 
Newtonian~\cite{Klement2019} variants of ECMC.
In addition, the pressure estimators of \subsect{subsec:PressureComputations}
are implemented (see \tab{tab:UltraNaive}). In detail, we provide pressure 
estimates from the wall density (using fit of the histogram), from the wall 
rifts using EDMD, and the wall rifts using ECMC, the latter testing the bias 
factor $N$ that is introduced because ECMC only moves a single disk. Moreover, 
we check our rift-average estimators for EDMD and for straight ECMC (that again 
differ by different biasing factors and mean values of perpendicular velocity 
components). Finally, we provide a test of the traditional fitting formula 
involving the pair-correlation function. All these estimators are of 
thermodynamic origin. As discussed in the main text, the kinematic estimators, 
including the virial formula, lead to identical formulas and need not be 
tested independently.

\subsubsection{Naive periodic-box programs}
\label{subsec:NaivePeriodicDisks}

The naive periodic-box programs contained in  the \HD package differ from
the \ultranaive programs only in that the number $N$ of disks and the radius
$\sigma$ can be set freely, and that the box is periodic.
These programs have some use for demonstration purposes, and to test the
more efficient algorithms for relatively small values of $N$.
Again, the Metropolis
algorithm, EDMD, and the four variants of ECMC are
implemented.
Run start from  crystalline initial
configurations. Configurations are output at fixed time 
intervals. EDMD and straight ECMC 
also output estimates of the pressure.

\subsubsection{State-of-the-art hard-disk programs}
\label{subsec:StateOfTheArtDisks}

The \HD package contains an optimized C++ code for 
straight ECMC, that is derived from the Fortran90 code used 
in~\cite{Bernard2011}. The GPU-based MPMC 
Cuda code used in this work derives from a general MPMC code for soft-sphere 
models and will be published elsewhere~\cite{Nishikawa2022}.
Pressures obtained from these implementations agree within very tight error
bars (see \tab{tab:StateOfArt}).
A Python script contained in the package analyzes
samples that were saved from
these two codes in the HDF5~\cite{hdf5} file format. It computes for example 
the global orientational order parameters $\Psi_6$.

\bibliography{General.bib,Historic.bib}
\end{document}